%% file: bare_jrnl.tex
\documentclass[journal]{IEEEtran}
\ifCLASSINFOpdf
   \usepackage[pdftex]{graphicx}
   \graphicspath{{../pdf/}{../jpeg/}}
   \DeclareGraphicsExtensions{.pdf,.jpeg,.png}
\else
\fi
\usepackage{amsmath}
\usepackage{tabularx}
\usepackage{enumitem}
\usepackage{multirow}
\usepackage{amssymb}
\usepackage{bm}
\usepackage{cite}

\begin{document}
\title{Review Polarity-wise Recommender}

\author{Han~Liu,~%\IEEEmembership{Member,~IEEE,}
        Yangyang~Guo,~
        Jianhua~Yin,~\IEEEmembership{Member,~IEEE,}
        Zan~Gao,
        and~Liqiang~Nie,~\IEEEmembership{Senior~Member,~IEEE}% <-this % stops a space
\thanks{Han Liu, Yangyang Guo, Jianhua Yin, and Liqiang Nie are with School of Computer Science and Technology, Shandong University, Qingdao 266200, China. E-mail: hanliu.sdu@gmail.com, guoyang.eric@gmail.com, jhyin@sdu.edu.cn, nieliqiang@gmail.com.}% <-this % stops a space
\thanks{Zan Gao is with Shandong Artificial Intelligence Institute, China. E-mail: zangaonsh4522@gmail.com.}% <-this % stops a space
\thanks{Liqiang Nie is the corresponding author.}
% \thanks{Manuscript received April 19, 2005; revised August 26, 2015.}
}


\markboth{IEEE Transactions on Neural Networks and Learning Systems}%
{Shell \MakeLowercase{\textit{et al.}}: Bare Demo of IEEEtran.cls for IEEE Journals}



\maketitle

\begin{abstract}
The \textit{de facto} review-involved recommender systems, utilizing review information to enhance recommendation, have received increasing interest over the past years.  Thereinto, one advanced branch is to extract salient aspects from textual reviews (i.e., the item attributes that users express) and combine them with the matrix factorization technique. However, existing approaches all ignore the fact that semantically different reviews often include opposite aspect information. In particular, positive reviews usually express aspects that users prefer, while the negative ones describe aspects that users dislike. As a result, it may mislead the recommender systems into making incorrect decisions pertaining to user preference modeling. Towards this end, in this paper, we present a Review Polarity-wise Recommender model, dubbed as RPR, to discriminately treat reviews with different polarities. To be specific, in this model, positive and negative reviews are separately gathered and utilized to model the user-preferred and user-rejected aspects, respectively. Besides, in order to overcome the imbalance of semantically different reviews, we further develop an aspect-aware importance weighting strategy to align the aspect importance for these two kinds of reviews. Extensive experiments conducted on eight benchmark datasets have demonstrated the superiority of our model as compared to several state-of-the-art review-involved baselines. Moreover, our method can provide certain explanations to the real-world rating prediction scenarios.
\end{abstract}

\begin{IEEEkeywords}
Review-involved Recommendation, Aspect-aware Recommendation, Review Polarity, Review Imbalance Problem.
\end{IEEEkeywords}

% \fi
\IEEEpeerreviewmaketitle

\section{Introduction}
\IEEEPARstart{N}{owadays}, posting reviews on e-commerce platforms has become ubiquitous among online shoppers
%a popular fashion for users
to share their purchasing experiences. These textual reviews usually contain rich semantic information about user preferences and item attributes, thereby playing an increasingly important role in recommender systems~\cite{cheng2019mmalfm}.
One typical benefit is that reviews enable the machine to effectively exploit more side information, and receive superior performance as compared with canonical matrix factorization-based methods \cite{liu2021factor}, as the latter methods utilize only the sparse rating matrix.

\begin{figure}[tbp]
  \centering
  \includegraphics[width=0.9\linewidth]{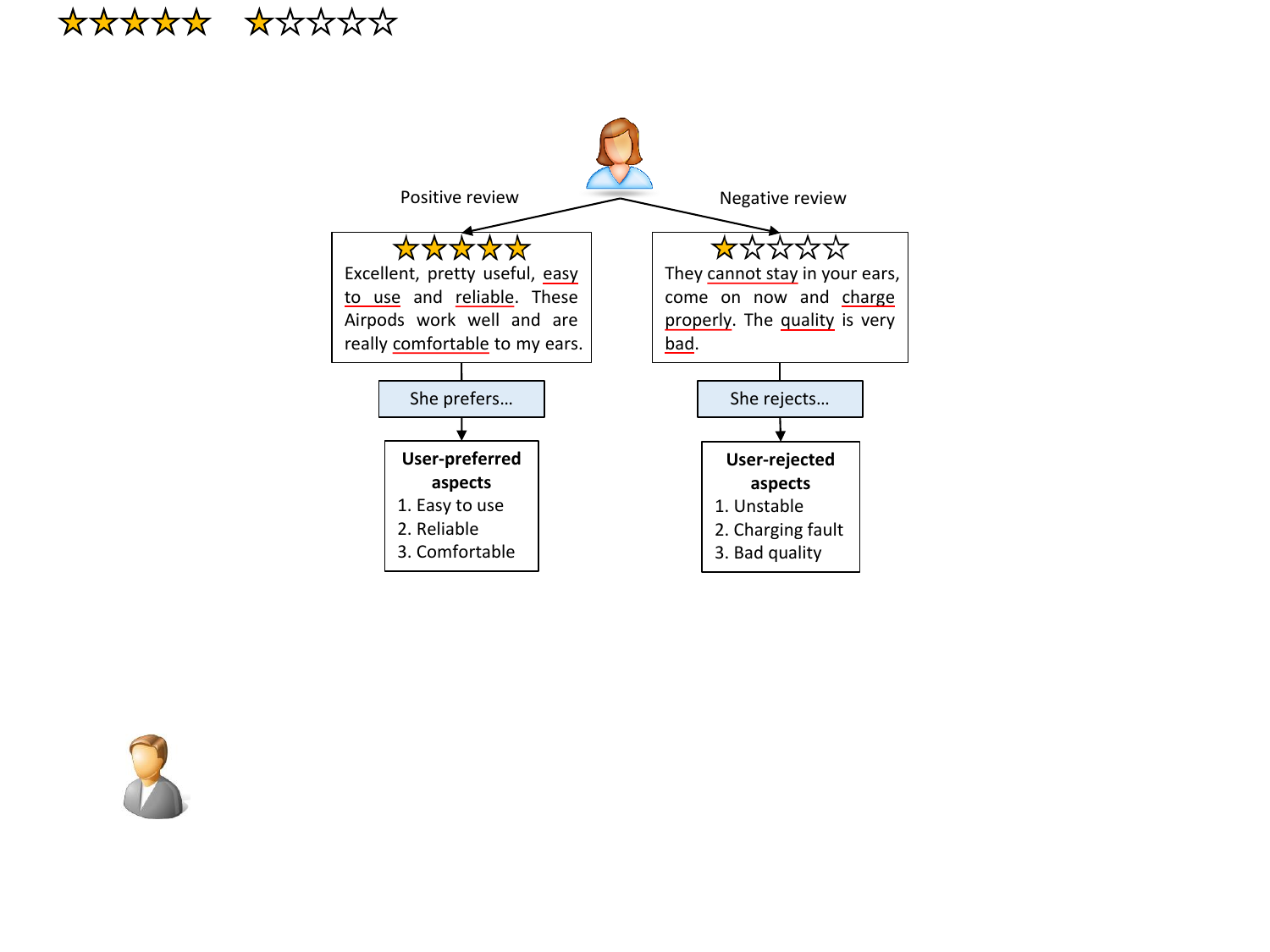}
%   \vspace{-0.5cm}
  \caption{An example of the opposite aspect information expressed in semantically different reviews.}
  \label{fig:introduction}
  \vspace{-0.5cm}
\end{figure}

Previous studies on review-involved recommendation mostly adopt a standard scheme:
% are mostly based on a hypothesis of two sides: 1) The user representation can be built from the user document (i.e., all the reviews that this user has posted); and 2) the item representation can be learned from the item document (i.e., all the reviews for this item).
% Based on this hypothesis, a typical recommendation scheme is developed:
the user and item documents are firstly constructed by merging the associated reviews (i.e., reviews of the user and reviews for the item), wherein each textual token is vectorized via word embedding methods~\cite{mikolov2013distributed,pennington2014glove}.
% And the word embedding methods
% \cite{mikolov2013distributed, pennington2014glove} are then adopted to map each review word into a dense representation.
The two types of documents are respectively processed via convolutional neural networks to generate the user and item representations, followed by a matching function (e.g., dot product and Factorization Machines~\cite{guo2020enhancing}) to predict the final rating score.
%This paradigm has proved to be highly effective,
Based on this scheme, methods like DeepCoNN \cite{zheng2017joint}, TransNets \cite{catherine2017transnets}, D-Attn \cite{seo2017interpretable}, and MPCN \cite{tay2018multi}  have achieved some improvements over other baselines.
Distinct from these approaches, recent efforts have been dedicated to the review aspect modeling~\cite{bauman2017aspect, wang2018explainable, cheng2018aspect, chin2018anr}. Foremost, the aspect is defined as follows:
\begin{itemize}[leftmargin=*]
\item \textbf{Aspect}~-~It is embodied with high-level semantics,  representing the attributes of items that users comment on in their reviews~\cite{cheng2018aspect}. For example, in the review `` Excellent, pretty useful, \textit{easy to use} and \textit{reliable}. These Airpods work well and are \textit{comfortable} to my ears'', the user mentioned the aspects \textit{easy to use}, \textit{reliable}, and \textit{comfortable} of item \textit{Airpods} (as shown in Fig.~\ref{fig:introduction}).
\end{itemize}
In general, these high-level aspect features are firstly extracted through well-developed tools, such as topic modeling~\cite{cheng2018aspect}, which are then integrated with the matrix factorization backbones~\cite{liu2021factor}.
% Different from the prior methods that process reviews to represent users and items, the recent efforts extract the aspect-aware information about users and items from reviews. For instance, ALFM~\cite{cheng2018aspect} utilizes reviews to model users' preferences and items' features on various aspects for finer-grained rating prediction.

\begin{figure}[tbp]
\centering
\begin{minipage}[t]{0.5\linewidth}
\centering
\includegraphics[width=\linewidth]{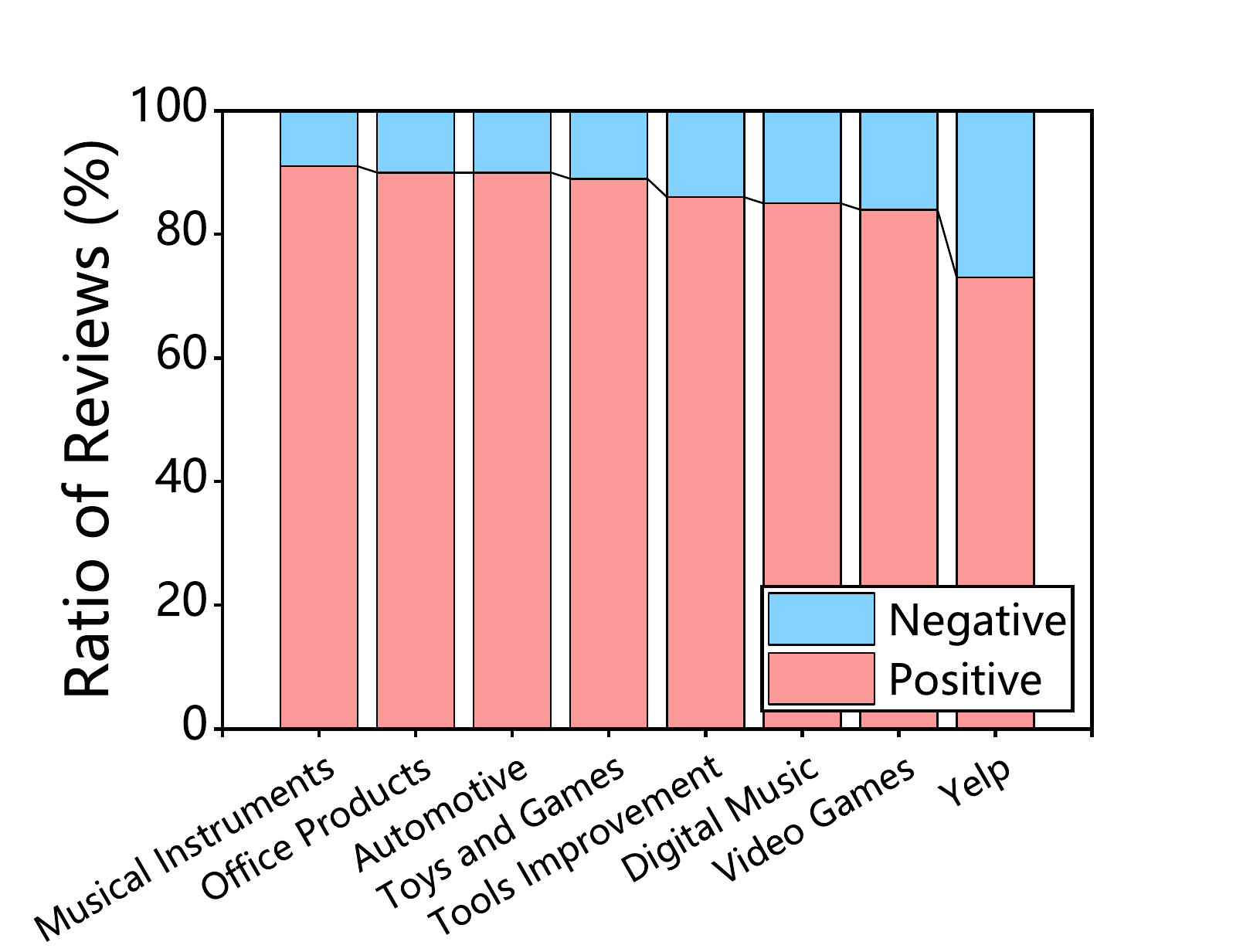}
\end{minipage}
\hspace{-0.8cm}
\begin{minipage}[t]{0.5\linewidth}
\centering
\includegraphics[width=\linewidth]{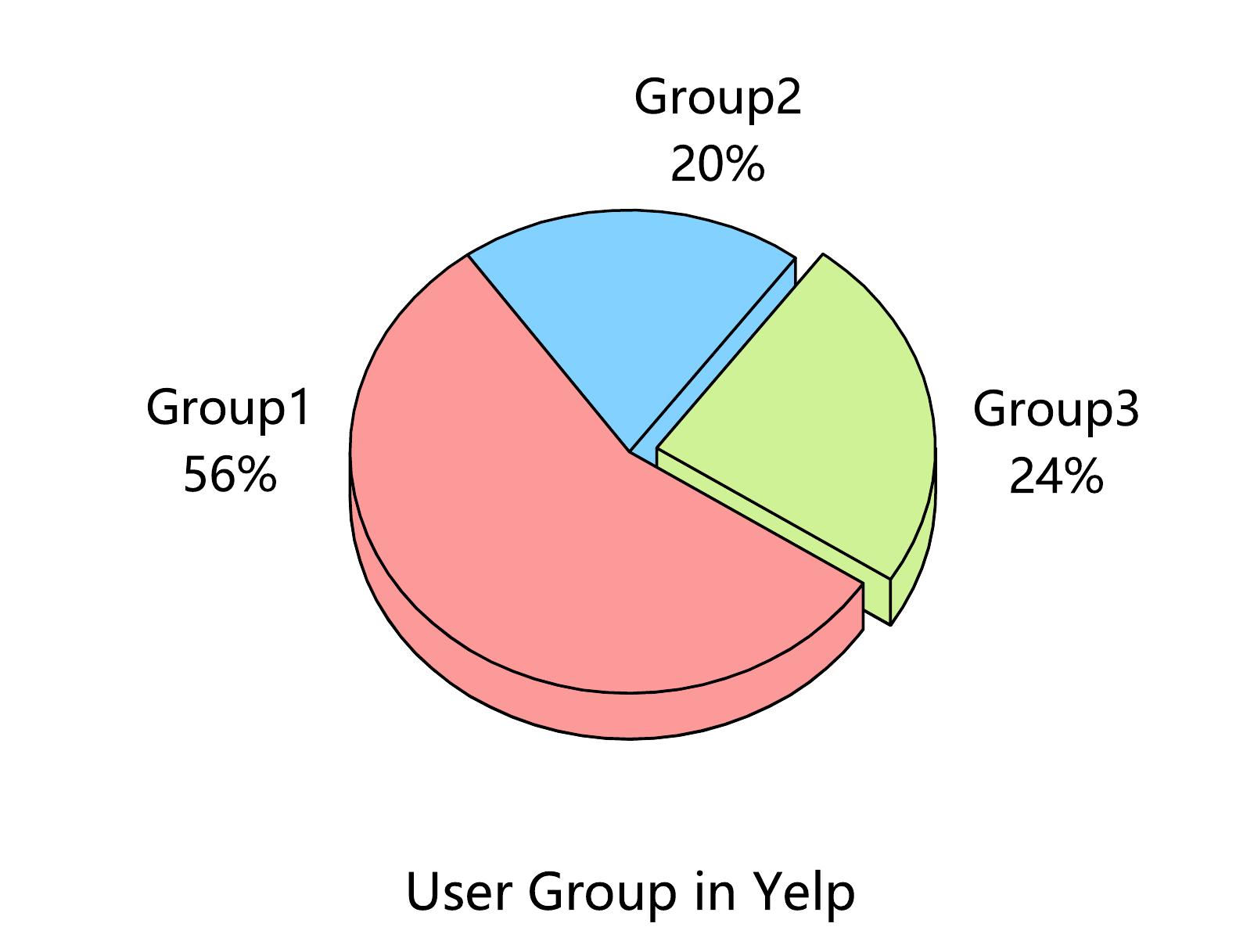}
\end{minipage}%
\centering
%\vspace{-0.4cm}
\caption{Imbalance illustration of the two-polarity reviews. The left subfigure shows the ratios of the two kinds of reviews in eight datasets. The right one shows various groups of users in Yelp. Group1 and Group2 respectively denote the numbers of users with positive and negative review ratios over $90\%$, and Group3 implies the remaining users.}
\label{fig:imbalance}
% \vspace{-0.5cm}
\end{figure}

Despite their notable progress, one issue hurts the performance of the existing review-involved methods is that the review polarities are not explicitly discriminated, i.e., all reviews are taken as positive feedback. In fact, users tend to convey their sentiments in reviews, i.e., higher rating scores often go with positive reviews, while negative ones meet lower scores frequently~\cite{chen2020try, lyu2021reliable, wang2017location}. Moreover, reviews with different polarities usually contain opposite aspect information.
Fig.~\ref{fig:introduction} shows two opposite polarities of reviews for \textit{bluetooth headsets}. The left one implies a positive review, describing aspects of an item that the user prefers, e.g., \emph{easy to use}.
On the contrary, the right one is a negative review that reveals unsatisfactory aspects of an item the user dislikes, e.g., \emph{unstable}. As illustrated  in Fig.~\ref{fig:imbalance},  negative reviews are quite common in existing datasets\footnote{http://jmcauley.ucsd.edu/data/amazon/.\label{footnote_amazon_data}}\,\footnote{https://www.yelp.com/dataset/challenge.\label{footnote_yelp_data}}. Simply integrating these two kinds of reviews as positive will mislead the models to recommend items similar to the ones with negative reviews in the future. It thus results in sub-optimal performance and deteriorates the user experiences and faithfulness to the platform. In the light of this, towards making the recommendation more personalized and convincing, we aim to distinguish the user-preferred aspects from the user-rejected ones via explicitly performing polarity discrimination. Nevertheless, discriminately treating positive and negative reviews
is non-trivial in recommendation, due to the following facts:
1) It is difficult to effectively exploit the semantic  information of each review word for aspect modeling.
2) How to accurately model the user preferences on both user-preferred and user-rejected aspects poses another challenge for us.
And 3) there exists severe imbalance regarding different review polarities in current e-commerce datasets, as shown in Fig.~\ref{fig:imbalance}.
For instance, some users tend to post much more positive reviews than negative ones.
It therefore exerts adverse impact on the extraction of aspect information,
degrading the recommendation performance.

\begin{figure*}[!tbp]
  \centering
  \includegraphics[width=0.9\linewidth]{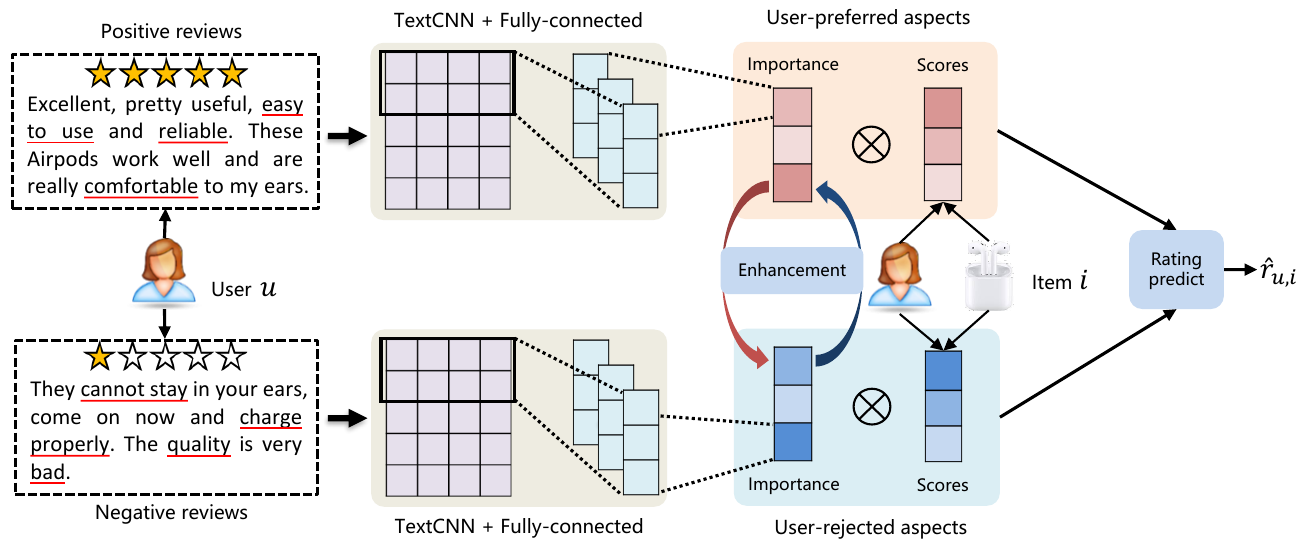}
%   \vspace{-0.3cm}
  \caption{The architecture of our proposed RPR model. From left to right, 1) two parallel convolutional networks are adopted to extract the importance of user-preferred and user-rejected aspects for user $u$ from $u$'s positive and negative reviews, respectively. Moreover, to overcome the imbalance problem between positive and negative reviews, an aspect-aware importance weighting module is appended to align the importance between the two polarities of reviews. 2) The user-preferred and user-rejected aspect scores of user $u$ towards item $i$ are separately learned. 3) The inner product of the score and importance vectors on the user-preferred aspects is calculated as the positive score of $u$ towards $i$, and then the negative score is similarly computed on the user-rejected aspects. RPR finally employs the subtraction of the negative score from the positive one to estimate the final rating.}
  \label{fig:architecture}
%   \vspace{-0.4cm}
\end{figure*}

In order to tackle the aforementioned issues, we present a Review Polarity-wise Recommender model, RPR for short, as shown in Fig.~\ref{fig:architecture}, to perceive the review polarity towards review-involved recommendation. In particular, for user $u$ and item $i$, we first leverage their latent factor embeddings to estimate two relevance score vectors for both the user-preferred and user-rejected aspects, whereby each element expresses the preference degree on the corresponding aspect\footnote{In RPR model, the aspects are implicit, and the number of aspects are fixed for all users.}. Meanwhile, traditional topic modeling has shown certain limitations in leveraging the abundant semantic information. We thus turn to a TextCNN in parallel on the review sentences. This module is expected to extract aspect importance as well as provide some explicit interpretations to the aspect modeling. Finally, the overall rating $r_{u,i}$ is estimated by subtracting the inner product of the score and importance vectors on the user-rejected aspects from the ones on the user-preferred aspects. Based on this, the positive and negative user preferences are seamlessly integrated to implement our RPR.

Besides, to overcome the imbalance between positive and negative reviews, we introduce an aspect-aware importance weighting module to capture the mapping relationship between the user-preferred and user-rejected aspect importance. The assumption is that if users focus on certain aspects they prefer, they will consistently pay roughly equal attention to the corresponding user-rejected ones, and vice versa. In view of this, according to the correspondence between user-preferred and user-rejected aspects,
RPR constructs a user-rejected aspect importance offset from the user-preferred aspect importance, which is further added to the original user-rejected one.
In this way, the two types of reviews mutually enhance each other for obtaining the aspect importance.

Overall, the main contributions of this paper are summarized in three-fold:
\begin{itemize}[leftmargin=*]
\item We propose a novel recommendation method to extract the semantic information of user-preferred  and user-rejected aspects
from positive and negative reviews, respectively.
To the best of our knowledge, this work is among the first efforts to treat reviews with different polarities discriminately in review-involved recommendation.
\item We devise an aspect-aware importance weighting component to construct the semantic mappings between user-preferred and user-rejected aspects.
%importance, the two types of aspect importance can complement with each other.
This design has been proven to be quite effective in solving the problem of data imbalance between positive and
negative reviews.
\item We conduct comprehensive experiments on eight benchmark datasets to evaluate the effectiveness of the proposed model. Extensive results demonstrate the state-of-the-art performance of RPR.
As a side contribution, we have released the codes, data, and parameters to facilitate researchers in this field\footnote{https://github.com/hanliu95/RPR.}.
\end{itemize}

The rest of this paper is organized as follows. Section 2 briefly reviews representative literature from two directions that are highly relevant to our work. Section 3 outlines RPR and its architecture, and describes how to optimize RPR. In Section 4 and 5, we experimentally evaluate RPR and analyze the evaluation results, respectively. We summarize the contributions and figure out the potential future research directions in Section 6.

%\IEEEpubidadjcol
\input{2_related_work}

\input{3_model}

\input{4_experiment}
\input{5_result}

\input{6_conclusion}

\ifCLASSOPTIONcaptionsoff
  \newpage
\fi



\bibliographystyle{IEEEtran}
\bibliography{ref}
\end{document}

%% file: 2_related_work.tex
\section{Related Work}
% In this section, we briefly review
% %related work in two different areas
% related literature from two directions
% which are highly relevant to our work: review-involved recommendation and deep learning in recommendation.%, and data imbalance.

\subsection{Review-involved Recommendation}
Recently, exploiting reviews to enhance the recommendation performance and interpretability
has been extensively studied in literature~\cite{seo2017interpretable, purushotham2012collaborative}.
%discussed and verified in many works.
%This not only improves the performance of recommendation but also provides a richer semantic modeling of user and item
%characteristics~\cite{zhang2014users}.
%In earlier days, many efforts have been mainly concentrated on extracting semantic features from reviews with
%the topic modeling and the language modeling approaches (yin).
%These relative works integrate the latent semantic topis into the latent factor models (yin).
%TLFM
Along this line, existing methods can be broadly classified into two categories. The first category mainly focuses on the user and item modeling from their separately corresponding documents.
%with deep learning techniques.
To be more specific, the user and item documents are firstly constructed by concatenating the user-posted and item-received reviews, respectively. And the word embedding techniques are then adopted to embed the textual document into a semantic matrix. Normally, a Convolutional Neural Network (CNN)~\cite{kim2016convolutional} is utilized to extract the user and item representations from the matrices. Thereafter, a matching function (e.g., dot product, Factorization Machines) can be employed to model the user-item interactions. Methods like DeepCoNN~\cite{zheng2017joint}, TransNets~\cite{catherine2017transnets}, D-Attn~\cite{seo2017interpretable}, and MPCN~\cite{tay2018multi} all adopt the above scheme, obtaining superior performance compared with the prior matrix factorization ones. 
% the outputs of which are fed into an elaborate network structure to learn embedding representations of the
% corresponding user and item.
% DeepCoNN employs parallel CNN networks to extract user and item representations from user and item documents \cite{zheng2017joint}.
% The extracted embeddings are concatenated and then fed into the Factorization Machines for rating prediction.
% TransNets \cite{catherine2017transnets} augments a DeepCoNN-like neural network by developing an additional multi-task learning scheme.
% More specifically, it learns an additional transform layer to infer the representation of the target user-item
% review, which is unavailable for the rating prediction.
% MPCN \cite{tay2018multi} argues that we should dynamically extract important reviews from user and item reviews depending on
% the current user-item target.
% It exploits a pointer-based co-attention mechanism to enable a multi-hierarchical information selection.
% Therefore, MPCN can identify both important reviews and words for the objective user and item.

%The works in another category of review-based recommendation
The second category aims to effectively learn the aspects from reviews for recommendation, namely, the aspect-aware recommender systems~\cite{chin2018anr,chen2016learning, zhang2014explicit}.
These methods can be further summarized into the following two groups.
The first group tries to extract aspects based on the existing NLP tools for sentiment analysis~\cite{zhang2014explicit, otter2020survey, he2015trirank}. The obtained aspect representation is then incorporated into a matrix factorization framework for more accurate recommendation.
For example, EFM \cite{zhang2014explicit} and MTER~\cite{wang2018explainable} resort to a phrase-level NLP tool for the aspect-level sentiment extraction. The second group devises specific internal components, e.g., topic modeling, to automatically learn explainable aspect representations of users and items from reviews~\cite{cheng2018aspect,chin2018anr,li2019capsule}. In particular, these internal components are leveraged to achieve the aspect-aware extraction of the semantic information in reviews. In a nutshell, compared with the first category, the aspect-aware methods are capable of extracting high-level semantic features  from reviews, delivering improved results as well as interpretability.

\subsection{Deep Learning in Recommendation}
A prevailing trend in recent years has been leveraging deep learning for recommendation. It is now extensively recognized that deep learning techniques are capable of modeling the complex and non-linear user-item interactions~\cite{luo2015nonnegative,deng2016deep}.
Generally speaking, the success of deep learning in recommendation mainly comes from two aspects: representation learning and matching function modeling~\cite{wang2015collaborative,GRCN,CLCRec}. Regarding representation learning, the entity embeddings, i.e., users and items, can be greatly enhanced with advanced tools in deep learning. For example, CNNs are used to enrich the item representation learning from both texts and images~\cite{zheng2017joint}, and RNNs show considerable advantage in session-based recommendation~\cite{han2019adaptive}.
%~\cite{he2017neural, he2016fast, mcauley2015image}. 
For the other aspect of matching function modeling, deep learning methods often utilize multi-level neural networks as the interaction function to effectively aggregate the low-level signals. Among these methods, Multi-Layer Perceptron (MLP) is distinguished with its strong capability to learn both the second-order and higher-order feature interactions~\cite{he2017NCF}.

Notably, the attention mechanism~\cite{zhu2019redundancy,xu2020cross,guo2019attentive} has been extensively integrated
into recommendation, which demonstrates 
promising results and great potentials in existing studies~\cite{huang2020efficient, guan2019attentive}. For example, ACF~\cite{chen2017attentive} introduces a hybrid  item- and component-level attention model. Meanwhile, NAIS~\cite{he2018nais} presents a neural attentive item similarity model for item-based collaborative filtering, enabling itself to identify the more important historical items in a user profile for rating prediction. In addition, AFM~\cite{xiao2017attentional} learns the weights of feature interactions in factorization machines via the neural attention networks. A\^{}3NCF~\cite{cheng20183ncf} introduces a topic model-based attention method, where the attention module is used to capture the attentive user preferences on each aspect of the target item.

%% file: 3_model.tex
\section{Proposed Method}
\subsection{Preliminaries}
In this subsection, we first briefly present the general framework for the aspect-aware model which exploits reviews to predict user-item interaction ratings, and point out its limitation caused by ignoring the review polarity. We then recapitulate how our RPR model can overcome the issue step-by-step.

Aspect-aware recommender systems assume that a user-item rating $r_{u,i}$ depends on user $u$'s score towards item $i$ on each aspect $a$ (i.e., aspect score $s_{u,i,a}$) and the importance of each aspect to $u$ (i.e., aspect importance $\rho_{u,a}$). It is worth noting that the aspects are implicitly defined throughout this paper following~\cite{cheng2018aspect,li2019capsule}. In general, the overall rating $r_{u,i}$ can be predicted by,
\begin{equation}
\label{eq:objective}
\hat{r}_{u,i}=\bm{\rho}_u^{\top}\mathbf{s}_{u,i}~,
\end{equation}
where the aspect importance $\bm{\rho}_u\in\mathbb{R}^{|\mathcal{A}|}$ is estimated based upon user reviews ($\mathcal{A}$ denotes the set of aspects, e.g., \{\textit{easy to use}, \textit{reliable}, \textit{comfortable}\} for \textit{Airpods}), and the aspect score $\mathbf{s}_{u,i}\in\mathbb{R}^{|\mathcal{A}|}$ is computed through matrix factorization relying on user-item interactions. However, the expression capability of these models is largely limited, since they treat all aspects as user-preferred and ignore the fact that reviews can contain negative opinions. This problem may lead to sub-optimal model performance, degrading the faithfulness of these methods.

To overcome the above issue, in this paper, we aim to predict $r_{u,i}$ via differently handling the two opposite polarities of reviews. One straightforward solution is to divide the aspects into user-preferred and user-rejected aspects from positive and negative reviews, respectively. The objective is then formulated as follows:
\begin{equation}
\label{eq:objective}
\hat{r}_{u,i}={\bm{\rho}^{p}_u}^{\top}\mathbf{s}^{p}_{u,i}
-
{\bm{\rho}^{r}_u}^{\top}\mathbf{s}^{r}_{u,i}~,
\end{equation}
where the subtraction of the two scores considers both user-preferred and user-rejected aspects during the rating prediction of $u$ towards $i$. Vector $\mathbf{s}^{p}_{u,i}$ ($\mathbf{s}^{r}_{u,i}$) consists of the estimated scores of $u$ towards $i$ on the user-preferred (user-rejected) aspects. $\bm{\rho}^p_u$ ($\bm{\rho}^r_u$) denotes the importance vector of the user-preferred (user-rejected) aspects for $u$, extracted from $u$'s positive (negative) reviews.

% Intuitively, a user $u$ rates an item $i$ after considering both user-preferred and user-rejected aspects. Accordingly, in this work, we aim to predict the rating of $u$ to $i$ by modeling $u$'s preferences towards $i$ on the two kinds of aspects. For addressing this task, we try to answer two questions: to what extent $u$ prefers (rejects) $i$ on the user-preferred (user-rejected) aspects, and the importance of these aspects for $u$. For the first question, we estimate the scores of $u$ towards $i$ on the user-preferred (user-rejected) aspects, denoted by a vector $\mathbf{s}^{p}_{u,i}$ ($\mathbf{s}^{r}_{u,i}$). For the second question, since semantically  different  reviews  usually  contain rich information about the opposite aspects,  we extract the importance of the user-preferred (user-rejected) aspects for $u$ by processing $u$'s positive (negative) reviews, denoted by an importance vector $\mathbf{\rho}^p_u$ ($\mathbf{\rho}^r_u$).

Moreover, in order to tackle the imbalance between positive and negative reviews, we intuitively assume that there exists a latent correlation of importance between the two kinds of aspects. We model the correlation to generate the aspect importance offsets $\bm{\mu}^p_u$ and $\bm{\mu}^r_u$ to enhance $\bm{\rho}^p_u$ and $\bm{\rho}^r_u$, respectively. Ultimately, the predictive model of RPR is given as:
\begin{equation}
\label{eq:objective}
\hat{r}_{u,i}={(\bm{\rho}^{p}_u+\bm{\mu}^p_u)}^{\top}\mathbf{s}^{p}_{u,i}
-
{(\bm{\rho}^{r}_u+\bm{\mu}^r_u)}^{\top}\mathbf{s}^{r}_{u,i}~.
\end{equation}

In the following, we will elaborate the proposed RPR model in three-fold: aspect score estimation, aspect importance extraction, and aspect importance offset learning.

\subsection{Aspect Score Estimation}
Following mainstream recommender models~\cite{he2017NCF}, we map users and items into a latent factor space and represent user $u$ and item $i$ by latent factor vectors $\mathbf{p}_u\in \mathbb{R}^f$ and $\mathbf{q}_i\in \mathbb{R}^f$, respectively.
According to~\cite{he2017NFM}, the interaction between $u$ and $i$ on each latent factor is characterized by $\mathbf{p}_u\odot\mathbf{q}_i$,
where $\odot$ represents the element-wise product between two vectors. Similar to~\cite{cheng2018aspect}, to leverage the latent factor-level interactions for aspect score prediction, we introduce two indicator matrices
$\mathbf{M}\in \mathbb{R}^{f\times |\mathcal{P}|}$ and $\mathbf{V}\in \mathbb{R}^{f\times |\mathcal{R}|}$ (where $|\mathcal{P}|$ and $|\mathcal{R}|$ are the numbers of user-preferred and user-rejected aspects, respectively),
to associate the latent factors to different user-preferred and user-rejected aspects, respectively.
The weight vector $\mathbf{m}_x$, which is the $x$-th column of $\mathbf{M}$, indicates which latent factor-level interactions
are related to the score of the $x$-th user-preferred aspect.
Similarly, the weight vector $\mathbf{v}_y$ indicates which interactions are related to the score of  the $y$-th user-rejected aspect.
Towards this end, we estimate the scores of the user-preferred and user-rejected aspects via the following formula,
\begin{equation}
\label{eq:score}
\left\{
\begin{split}
&\mathbf{s}^{p}_{u,i}=\mathbf{M}^{\top}(\mathbf{p}_u\odot\mathbf{q}_i)
\\
&\mathbf{s}^{r}_{u,i}=\mathbf{V}^{\top}(\mathbf{p}_u\odot\mathbf{q}_i)
\end{split}
\right.
,
\end{equation}
where $\mathbf{s}^{p}_{u,i}$ and $\mathbf{s}^{r}_{u,i}$ represent user's preference scores on the user-preferred and user-rejected aspects of items,  respectively.

\begin{figure}[tbp]
  \centering
  \includegraphics[width=\linewidth]{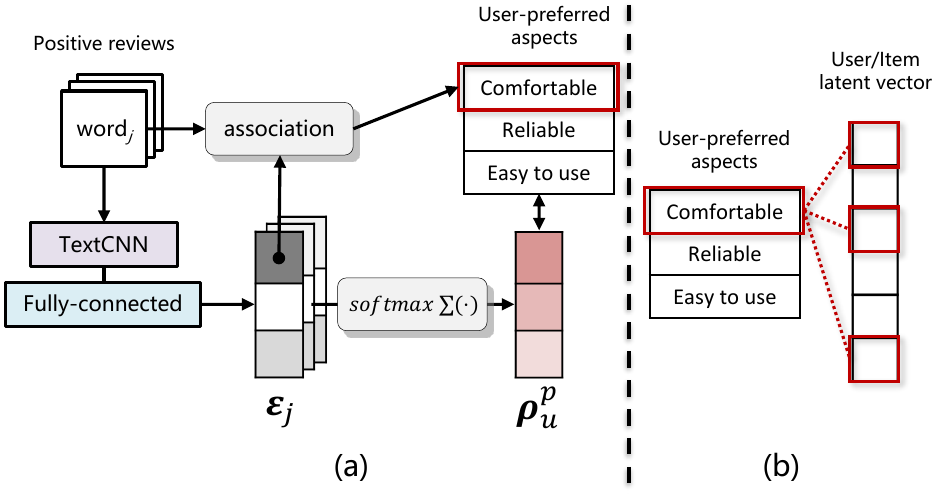}
  \vspace{-0.5cm}
  \caption{Aspect importance modeling based upon the review words and correlation between aspects and latent vectors. (a) Aspect importance modeling. Each review word is embedded with TextCNN and Fully-connected layers to produce semantic embedding $\bm{\varepsilon}$. The user-preferred aspect importance $\bm{\rho}^p_u$ is then estimated via the summation of all the word semantic embeddings followed by a $softmax$ function. We then associate the largest element $\varepsilon_{j,1}$ of $\bm{\varepsilon}_j$ to its corresponding aspect, i.e., the first aspect is expressed with the $j$-th word \emph{comfortable}. And (b) an exemplar correlation between aspect and latent vectors. The three elements circled with red boxes in the latent vector are related to the first aspect \emph{comfortable}.}
  \label{fig:aspect2latent-vector}
%   \vspace{-0.5cm}
\end{figure}

\subsection{Aspect Importance Extraction}
We define that the user $u$'s positive document $D_{u}^{pos}$ is constructed by collecting all the positive reviews posted by $u$, and the user's negative document $D_{u}^{neg}$ is built in a similar manner.
It is widely accepted that users tend to comment on aspects with opposite attitudes pertaining to different polarities of reviews. In addition, users would individually care more about certain aspects than others. For example, fashion enthusiasts often focus on the user-preferred aspect ``\textit{fashionable style}'' of ``\textit{clothing}'' items, as the review words like ``\textit{fashion sense}'' frequently appear in their reviews. In the following, we mainly detail the formulation on how to extract the user-preferred aspect importance by word-wise extraction of $D^{pos}_u$, while the user-rejected one can be obtained in a similar way.

%By means of pre-trained word embedding models, we first project each word in $D_{u}^{pos}$ to its embedding representation:
%$\mathbf{D}_{u}^{pos}=[\mathbf{e}_1, ..., \mathbf{e}_l]$, where $l$ is the number of words in the document,
Firstly, we use the pre-trained word embeddings to initialize the word representations in the positive document,
where $\mathbf{e}_j\in \mathbb{R}^d$ is the embedding vector for the $j$-th word, and $d$ denotes the embedding dimension.
We then adopt a CNN model to extract the contextual information of each word~\cite{conneau2017very, kim2014convolutional}.
The convolution layer can be regarded as a tensor $\mathbf{K}$, which consists of $N$ neurons. Each neuron uses a filter with size $d\times c$ spanning $\epsilon=\frac{c-1}{2}$ words on both sides of each word.
The latent contextual feature vector for the $j$-th word is formulated as follows,
\begin{equation}
\label{eq:cnn}
\mathbf{c}_j = ReLU(\mathbf{W}_c([\mathbf{e}_{j-\epsilon}\cdots\mathbf{e}_j\cdots\mathbf{e}_{j+\epsilon}]\ast\mathbf{K})+\mathbf{b}_c),
\end{equation}
where $\mathbf{c}_j\in\mathbb{R}^N$ is the latent contextual feature vector for the $j$-th word, $\mathbf{W}_c\in\mathbb{R}^{N\times N}$ and $\mathbf{b}_c\in\mathbb{R}^{N}$ denote
the weight matrix and bias vector, respectively.

Secondly, we focus on how to extract the user-preferred aspect importance based on the word contextual feature vectors.
% Intuitively, a user always emphasizes which good aspects of items he/she likes and what extent in his/her positive reviews.
% Thus the latent contextual feature of each word contributes importance to different good aspects.
In fact, some salient words are supposed to contribute more to the aspect importance modeling of users.
For example, the word ``cost-effective'' in reviews implies that the user probably puts more emphasis on aspects like ``good price'' and ``effectiveness''.
In addition, it is natural that if an aspect-specific word is frequently mentioned in $D_{u}^{pos}$,
the user $u$ will attach more importance to  the aspects related to this word.
For example, ``delicious'' and ``yummy'' would be repeatedly written by a user who pays much attention to the ``good taste'' aspect.
Inspired by this, we develop a fine-grained semantic extraction network for the aspects. Specifically, we resort to the fully-connected layers to automatically discriminate the importance contribution of each review word,
\begin{equation}\label{eq:rho_p}
\left\{
\begin{split}
&\bm{\varepsilon}_j=ReLU(\mathbf{W}_{\rho}\mathbf{c}_j + \mathbf{b}_{\rho})\\
&\bm{\rho}^p_u=softmax(\sum\nolimits_{j=1}^{l_{pos}} \bm{\varepsilon}_j)
\end{split}
\right.
,
\end{equation}
where the semantic embedding $\bm{\varepsilon}_j\in \mathbb{R}^{|\mathcal{P}|}$ is projected from the $j$-th word contextual features through a matrix $\mathbf{W}_{\rho}\in\mathbb{R}^{|\mathcal{P}|\times N}$, bias $\mathbf{b}_{\rho}\in\mathbb{R}^{|\mathcal{P}|}$, and the $ReLU$ activation function. $l_{pos}$ represents the number of words in the user's positive document, and $\bm{\rho}^p_u\in \mathbb{R}^{|\mathcal{P}|}$ is the user-preferred aspect importance, reflecting the emphasis degree on each user-preferred aspect attached by user $u$.
% function, considering the word frequency.

From the above formulation, it can be seen that each element in the word semantic embedding contributes distinctively to the aspect importance modeling. In the light of this, we adopt a loose hypothesis that each aspect associates closely with certain review words. We hence assume that one word can express one aspect (i.e., one element in the aspect vector) if the corresponding element from the word semantic embedding is the largest, since the dimensions of word semantic embedding and aspects are the same\footnote{We can also use $k$ words to express an aspect with sorting the corresponding element values.}. For example, as shown in Fig.~\ref{fig:aspect2latent-vector}, the $j$-th word, whose first element $\varepsilon_{j,1}$ is the largest in the semantic embedding $\bm{\varepsilon}_j$, can be associated with the first aspect.

Based on the same procedure, we can also extract the user's importance vector for user-rejected aspects $\bm{\rho}^r_u$
with a similar module from her/his negative document $D_{u}^{neg}$.

\subsection{Aspect Importance Offset Learning}
Though we have obtained the expected scores and the corresponding importance distribution for the user-preferred and user-rejected
aspects, it is still insufficient to correctly predict the overall rating.
The reason is that in a practical scenario, it is common that the objective user $u$
provides much more reviews from one polarity than the other, referring to Fig.~\ref{fig:imbalance}.
%only wrote limited reviews so that his positive and negative documents are imbalanced.
For a better understanding,  we make an extreme assumption that if user $u$ only posts positive reviews,  the learned user-rejected aspect importance vector $\bm{\rho}^r_u$ would approximately approach zero in RPR.
Furthermore, if we leverage the user-rejected aspect importance vector to predict the ratings toward the items in aspects the user distastes,  the predicted rating would be  unfavorable.

In order to solve the imbalance in reviews with different polarities, we propose to construct the relations between the user-preferred and user-rejected aspect
importance. In particular, we intuitively assume that if a user attaches more importance to a user-preferred aspect,
she/he will correspondingly pose roughly equal importance to the related user-rejected aspects.
For example, a user pays attention to the user-preferred aspect of ``elegant environment'' and has chosen a restaurant, she/he would similarly care about the user-rejected aspects like ``obsolete decoration'' and ``poor sanitary situation''.
Based on this assumption, we leverage an aspect-aware importance weighting module to construct the mapping relationship between the user-preferred and user-rejected aspect importance.

In order to construct the relationships between two opposite polarities of aspects, we utilize the common \textit{user/item latent space} as a bridge. Given that the aspect is associated with specific latent factors as shown in Eqn.(4), the related aspects from the other polarity are expected to be more similar in the latent space. As weight vectors $\mathbf{m}_x$ and $\mathbf{v}_y$ are responsible for the latent aspect modeling (as mentioned in Section 3.2), RPR takes these indicator vectors as inputs to the aspect-aware importance weighting module for measuring the similarity among aspects. Specifically, the following function is employed to obtain the attention weight of the user-rejected aspect $y$ to each user-preferred
aspect,
\begin{equation}
\left\{
\begin{split}
&\phi'_{y,x}=\mathbf{h}_{a}^{\top} ReLU(\mathbf{W}_{a}(\mathbf{v}_y\odot\mathbf{m}_x)+\mathbf{b}_{a})
\\
&\phi_{y,x}=\frac{\exp(\phi'_{y,x})}{\sum_{x\in\mathcal{P}}\exp(\phi'_{y,x})}
\end{split}
\right.
,
\end{equation}
where $\phi_{y,x}$ denotes the attention weight of the user-rejected aspect $y$ to the $x$-th user-preferred aspect,
and $\mathbf{h}_{a}$, $\mathbf{W}_{a}$, and $\mathbf{b}_{a}$ are the parameters of the aspect attention network.
For simplicity, we concatenate these attention weights into a matrix
$\mathbf{\Phi}\in \mathbb{R}^{|\mathcal{P}|\times |\mathcal{R}|}$.
The $y$-th column $\bm{\phi}_y$ of matrix $\mathbf{\Phi}$ denotes the attention weight vector of the user-rejected aspect $y$, whose $x$-th element is $\phi_{y,x}$.
% In this way, we can leverage a similar attention network with different parameters to learn the attention weights for the user-preferred aspect $x$,
% \begin{equation}
% \left\{
% \begin{split}
% &\psi'_{x,j}=\mathbf{h}_{2}^T ReLU(\mathbf{W}_{2}(\mathbf{m}_x\odot\mathbf{v}_j)+\mathbf{b}_{2})
% \\
% &\psi_{x,j}=\frac{\exp(\psi'_{x,j})}{\sum_{j\in\mathcal{R}}\exp(\psi'_{x,j})},
% \end{split}
% \right.
% \end{equation}
% where $\psi_{x,j}$ denotes the attention weight of the user-preferred aspect $x$ to the $j$-th user-rejected aspect,
% and $\mathbf{h}_{2}$, $\mathbf{W}_{2}$, and $\mathbf{b}_{2}$ are the parameters of the aspect attention network. Similarly, we integrate these attention weights into a matrix
% $\mathbf{\Psi}\in \mathbb{R}^{|\mathcal{R}|\times |\mathcal{P}|}$.
% The $x$-th column $\bm{\psi}_x$ of matric $\mathbf{\Psi}$ denotes the attention weight vector of the user-preferred aspect $x$, whose $j$-th element is $\psi_{x,j}$.

\begin{figure}[tbp]
  \centering
  \includegraphics[width=\linewidth]{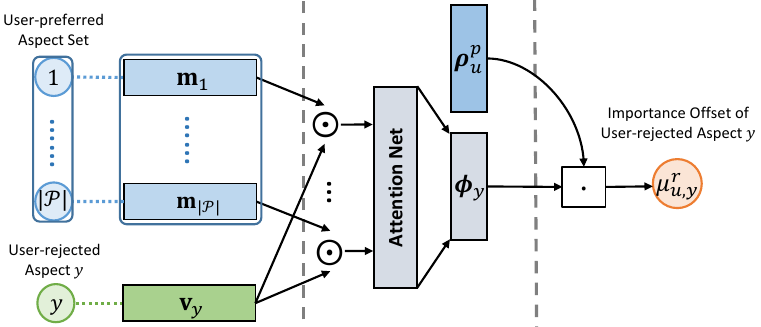}
  \caption{The generation process of importance offset for the $y$-th user-rejected aspect. The similarity between $\mathbf{v}_y$ and each column of $\mathbf{M}$ is computed via the attention net, which is then combined with the initial user-preferred aspect importance vector $\bm{\rho}^p_u$.}
  \label{fig:attention}
  \vspace{-0.45cm}
\end{figure}

We take the attention weights as the mapping relationships between the user-preferred and user-rejected aspect importance.
As for the aforementioned imbalance problem,
we can employ an offset vector as an addition to the user-rejected aspect importance vector by utilizing the matrix $\mathbf{\Phi}$.
Thus, the enhanced user-rejected aspect importance vector of user $u$ equals,
\begin{equation}
\label{eq:rhob}
\left\{
\begin{split}
&\bm{\mu}^r_u = \mathbf{\Phi}^{\top}\bm{\rho}^p_u
\\
&\bm{\rho}^{r+}_u = \bm{\rho}^r_u + \bm{\mu}^r_u
\end{split}
\right.
,
\end{equation}
where $\bm{\mu}^r_u$ denotes the offset for user-rejected aspect importance,
and $\bm{\rho}^p_u$ is the user-preferred aspect importance vector that has already been extracted in Eqn.($\ref{eq:rho_p}$).
The $y$-th element of the offset vector $\bm{\mu}^r_u$ equals the inner product $\bm{\phi}^{\top}_y\bm{\rho}^p_u$,
and the attention weights of the $y$-th user-rejected aspect on each user-preferred aspect model a linear relationship,
which maps the extracted user-preferred aspect importance to the objective user-rejected aspect importance space. In this way, we can achieve a more intuitive user-rejected aspect importance vector of a user from her/his positive reviews indirectly,
even if the negative reviews are inadequate. Fig.~\ref{fig:attention} illustrates the generation process of importance offset for the $y$-th user-rejected aspect. Similarly, we can obtain the enhanced user-preferred aspect importance $\bm{\rho}^{p+}_u$, where the process is omitted due to space limitation.
% Similarly, the attention weight matrix $\mathbf{\Psi}$ is employed to get the enhanced user-preferred aspect importance,
% \begin{equation}
% \label{eq:rhog}
% \left\{
% \begin{split}
% &\bm{\mu}^p_u = \mathbf{\Psi}^T\bm{\rho}^r_u,
% \\
% &\bm{\rho}^{p+}_u = \bm{\rho}^p_u + \bm{\mu}^p_u,
% \end{split}
% \right.
% \end{equation}
% where $\bm{\mu}^p_u$ denotes the offset for user-preferred aspect importance. Note that if the negative reviews are sparse, the user-rejected aspect importance would pose little influence to the enhanced user-preferred aspect importance modeling, which can be assured by the attention weight matrix.
Extensive experiments have demonstrated that the offset components can effectively resolve the imbalance
problem between the positive and negative reviews, which will be elaborated in Section 5.2.

\subsection{Overall Objective}
Up to now, we have obtained the scores and importance of the user-preferred and user-rejected aspects, respectively.
The expected rating $\hat{r}_{u,i}$ that user $u$ would give to item $i$ is computed as follows,
\begin{equation}
\label{eq:objective}
\hat{r}_{u,i}={\bm{\rho}^{p+}_u}^{\top}\mathbf{s}^{p}_{u,i}
-
{\bm{\rho}^{r+}_u}^{\top}\mathbf{s}^{r}_{u,i}~.
\end{equation}
% \begin{equation}
% \hat{r}_{u,i}=(\mathbf{p}_u\odot\mathbf{q}_i)^T\mathbf{M}\bm{\rho}^{g+}_u
% -
% (\mathbf{p}_u\odot\mathbf{q}_i)^T\mathbf{V}\bm{\rho}^{b+}_u.
% \end{equation}
% where $\bm{\cdot}$ is the inner product operation. The inner product combines the score and importance of the user on each user-preferred aspect of the item to compute the overall positive score, as well the overall negative score.
% $(\mathbf{p}_u\odot\mathbf{q}_i)^T\mathbf{V}$ computes the score vector of bad aspects, and makes inner product with the final bad aspect importance as the overall negative score.
% Ultimately, the overall rating is predicted by the difference between these two scores.
% where $\bm{\cdot}$ denotes the inner product of vectors.
In this way, both the user positive and negative preferences can be simultaneously considered and discriminated.
% Fig.~\ref{fig:architecture} illustrates how RPR predicts the overall rating of a user $u$ to an item $i$.
% Algorithm $1$ summarizes the rating prediction algorithm of our RPR model, and provides a clear indication of the time complexity of algorithm steps, which will be discussed in the experimental part.
The estimation of model parameters is to minimize the rating prediction error on the training dataset. In this way, the objective function is,
\begin{equation}
\label{eq:obj}
\begin{split}
\min_{\mathbf{p}_u, \mathbf{q}_i, \mathbf{M}, \mathbf{V}, \bm{\Theta}}
\frac{1}{2}\sum_{u,i}(r_{u,i}-\hat{r}_{u,i})^2
+\beta_1\big(\|\mathbf{M}\|_1 + \|\mathbf{V}\|_1\big)
\\
+\frac{\beta_2}{2}\Big(\|\mathbf{p}_u\|^2_2 + \|\mathbf{q}_i\|^2_2
+ \sum_{\bm{\theta}\in\mathbf{\Theta}}\|\bm{\theta}\|^2_2\Big).
\end{split}
\end{equation}
With the minimization of this objective function, all model parameters can be effectively updated through the gradient decent strategy. The involved parameters include  user and item latent vectors $\mathbf{p}_u$ and $\mathbf{q}_i$, indicator matrices $\mathbf{M}$ and $\mathbf{V}$ from the aspect score estimation module, and the remaining parameters $\mathbf{\Theta}$ from the aspect importance extraction and the aspect importance offset learning modules.
% To be more specific, $\mathbf{\Theta}$ contains the deep learning parameters of our model, such as the convolution tensor $\mathbf{K}$, the fully connected parameters $\mathbf{W}_{\rho}$ and $\mathbf{b}_{\rho}$, and the attention network parameter $\mathbf{h}_a$.
$\|\cdot\|_1$ and $\|\cdot\|_2$ respectively denote the $l_1$ and $l_2$ regularization norm, with the corresponding hyper-parameters $\beta_1$ and $\beta_2$.
Considering the fact that $l_1$ regularization yields sparse solution of the weights,
we thus use $l_1$ norm to obtain approximately binary matrices $\mathbf{M}$ and $\mathbf{V}$ for better indication ability.
The $l_2$ regularization of $\mathbf{p}_u$, $\mathbf{q}_i$, and $\bm{\theta}$ prevents these parameters from uncontrollable values and over-fitting.

\textbf{Optimization.}
%Since the parameters in our proposed model mainly consists of three parts:
%the user and the item latent factor matrices, the weight matrices $\mathbf{W}$ and $\mathbf{V}$,
%and the set of remaining parameters in the deep-learning modules (\emph{e.g.}, the convolutional network and the attention network),
%we apply alternating optimization strategy to learn the model parameters,
%taking turns to update each of the three parts.
We leverage the Adam optimization algorithm~\cite{kingma2014adam} to learn all the parameters by minimizing the objective function in Eqn.(\ref{eq:obj}).
Note that for training stability, the parameter is optimized in the following sequence:
the user matrix, the item matrix, the weight matrices, and the remaining parameters.

%% file: 4_experiment.tex
\section{Experimental Setup}
In this section, we first presented the evaluation datasets, and then introduced our experimental settings. Finally, we listed several baseline methods for comparison.
\subsection{Datasets}
%To verify the effectiveness of our proposed model, we apply seven datasets from two resource:
We conducted experiments on two publicly available datasets:
Amazon\textsuperscript{\ref{footnote_amazon_data}} and Yelp\textsuperscript{\ref{footnote_yelp_data}}.
The Amazon dataset provides rich review information with rating scores.
In our experiments, we applied its seven sub-datasets:
\emph{Musical Instruments, Office Products, Digital Music, Tools Improvement, Automotive, Toys and Games, and Video Games}.
%The datasets are preprocessed in a 5-core fashion (\emph{i.e.}, each user and item have at least 5 reviews to be included).
Yelp is a famous online review platform for business, such as restaurants, bars, and spas.
We selected the dataset from the latest version and used a 20-core setting to provide a denser dataset.
Each record in our datasets consists of user ID, item ID,  rating, and the corresponding review text.
For all datasets, we filtered out the empty-review records.
The target rating score used in these datasets ranges from 1 to 5.
Similar to~\cite{wang2021denoising}, reviews with rating scores  higher than or equal to 3 are split into positive documents, while the ones lower than 3 are regarded negative on all datasets.
Table~\ref{tab:statistics} summarizes the detailed statistics of the evaluated datasets, where ``pos$/$neg ratio'' denotes relative proportions between positive and negative reviews of each dataset.

\begin{table}[htbp]\centering
\caption{Statistics of the evaluated datasets.}
\vspace{-0.3cm}
\label{tab:statistics}
\begin{tabular}{ccccc}
    \hline
    \textbf{Datasets} & \textbf{Users} & \textbf{Items} & \textbf{Ratings} & \textbf{pos$/$neg ratio}
    \\
    \hline
    Musical Instruments & 1,429 & 900 & 10,262 & 10.11
    \\
    Office Products & 4,905 & 2,420 & 53,258 & 9.10
    \\
    Digital Music & 5,541 & 3,568 & 64,706 & 5.67
    \\
    Tools Improvement & 16,638 & 10,217 & 134,345 & 6.14
    \\
    Automotive & 2,928 & 1,835 & 20,473 & 9.20
    \\
    Toys and Games & 19,412 & 11,924 & 167,597 & 8.09
    \\
    Video Games & 24,303 & 10,672 & 231,577 & 5.25
    \\
    Yelp & 40,500 & 58,755 & 2,024,283 & 2.70
    \\
    \hline
\end{tabular}
\end{table}

For each dataset, we randomly split its records into two parts: $80\%$ for training and the rest $20\%$ for
testing.
Moreover, $10\%$ records in the training set are randomly selected as the validation set for hyper-parameter tuning.
Note that we slightly adjusted the training and testing sets to ensure that at least one record for each user/item would be included
in the training set.
The target reviews in the validation and testing sets are excluded since they are unavailable in the practical scenario.

\subsection{Experimental Settings}
\textbf{Evaluation Metrics.} To thoroughly evaluate our model and
the baselines, we adopted the \textbf{\textbf{MSE}} (Mean Squared Error) and the \textbf{MAE} (Mean Absolute Error) as the evaluation metrics to measure
the rating prediction performance.

% \textbf{MSE} computes the average squared difference between the estimated ratings and the actual ratings:
% \begin{equation*}
%     MSE=\frac{\sum_{u,i}^{\mathcal{U},\mathcal{I}}(r_{u,i}-\hat{r}_{u,i})^2}{\mathcal{N}_{test}},
% \end{equation*}
% where $\mathcal{N}_{test}$ is the number of ground-truth samples in the testing set.

% \textbf{MAE} is an arithmetic average of the absolute errors between the estimated ratings and the actual ratings:
% \begin{equation*}
%     MAE=\frac{\sum_{u,i}^{\mathcal{U},\mathcal{I}}|r_{u,i}-\hat{r}_{u,i}|}{\mathcal{N}_{test}}.
% \end{equation*}
\textbf{Implementation Details.} We implemented our model via the development tool Tensorflow\footnote{http://www.tensorflow.org.}.
The embedding matrix of the document is initialized via word vectors which have been pre-trained in GloVe\footnote{https://nlp.stanford.edu/projects/glove/.} (used 50-$d$ vectors for its efficiency).
For the convolutional layer in the proposed model, the number and size of filters are set to 50 and 3, respectively.
We utilized the popular approach of Xavier \cite{glorot2010understanding} to initialize the weights
%and bias
in our model.
And we adopted grid search to tune the hyper-parameters based on the results from the validation set. Moreover, we varied the number of both the user-preferred and user-rejected aspects within the set $\{1, 2, 3, 4, 5\}$,
the dimension of user and item latent factor vectors amongst $\{4, 8, 16, 32, 64\}$,
the learning rate amongst \{1E-05, 1E-04, 1E-03, 1E-02\},
and the size of training mini-batch amongst \{100, 200, 500, 1,000\}.
\subsection{Baseline Comparison}
We compared the performance of our proposed method with a series of state-of-the-art recommendation methods. To summarize, we divided the baselines into three categories. The first category is interaction-based, including: \textbf{MF}, \textbf{FM}~\cite{rendle2012factorization}, \textbf{MLP}~\cite{he2017NCF}, and \textbf{NeuMF}~\cite{he2017NCF}. The second category is plain review-involved, including: \textbf{DeepCoNN}~\cite{zheng2017joint}, \textbf{TransNet}~\cite{catherine2017transnets}, and \textbf{MPCN}~\cite{tay2018multi}. The last category is aspect-aware, including: \textbf{ALFM}~\cite{cheng2018aspect} and \textbf{CARP}~\cite{li2019capsule}.

We adopted the publicly available implementations of FM, NeuMF, DeepCoNN, TransNet, MPCN, ALFM, and CARP in our experiments.
When training all these baseline models, we set the maximum training epoch to $50$ for a fair comparison.
For the interaction-based models, we varied the embedding size within $\{8, 16, 32, 64\}$.
All word embeddings in review-based baselines are initialized using pre-trained word vectors in GloVe~\cite{pennington2014glove} or word2vec~\cite{mikolov2013distributed}\footnote{https://code.google.com/archive/p/word2vec/.}.
For the convolutional layer in the review-based models, the filter size is set to 3, and we tested various numbers of filters
amongst $\{20, 50, 100, 200\}$ to select the optimal one.
Dropout is appended after all fully-connected and convolution layers with a dropout rate of $0.2$.
For TransNet, we used two transform layers, following the model setting adopted in the original paper~\cite{catherine2017transnets}.
For MPCN, the number of pointers is tuned amongst $\{1, 3, 5, 8, 10\}$. For ALFM, we tuned the numbers of aspects and latent factors following the original paper~\cite{cheng2018aspect}. For CARP, we took the suggestion in the original paper~\cite{li2019capsule}, and set the capsule number and the predefined threshold value to 5 and 3, respectively.

\begin{table*}\centering
\caption{Performance comparison on eight datasets. The best performance is highlighted in boldface. $\Delta_{CA}$ denotes the relative improvement ($\%$) of RPR over the best baseline CARP. P-value reflects the t-test result of RPR compared with CARP.}
\label{tab:q1}
\vspace{-0.3cm}
\begin{tabular}{|c|p{1.3cm}<{\centering}|p{1.3cm}<{\centering}|p{1cm}<{\centering}|p{1cm}<{\centering}|p{1cm}<{\centering}|p{1cm}<{\centering}|p{1.2cm}<{\centering}|p{1.2cm}<{\centering}|}
    \hline
    \multirow{2}*{Datasets} & \multicolumn{2}{c|}{Musical Instruments} & \multicolumn{2}{c|}{Office Products} & \multicolumn{2}{c|}{Digital Music} & \multicolumn{2}{c|}{Tools Improvement}
    \\
    \cline{2-9}
    {} & MSE & MAE & MSE & MAE & MSE & MAE & MSE & MAE
    \\
    \hline
    MF& 1.970 & 1.404 & 1.144 & 1.070 & 1.956 & 1.399 & 1.560 & 1.249
    \\
    FM& 1.167 & 1.080 & 1.098 & 1.048 & 1.395 & 1.181 & 1.320 & 1.149
    \\
    MLP& 1.082 & 1.040 & 1.120 & 1.058 & 1.356 & 1.164 & 1.351 & 1.162
    \\
    NeuMF& 1.187 & 1.089 & 1.078 & 1.038 & 1.334 & 1.155 & 1.314 & 1.146
    \\
    \hline
    D-CON& 1.286 & 1.135 & 0.975 & 0.825 & 1.330 & 1.153 & 1.248 & 1.063
    \\
    T-NET& 1.130 & 0.872 & 0.951 & 0.789 & 1.325 & 1.052 & 1.124 & 0.923
    \\
    MPCN& 0.923 & 0.860 & 0.879 & 0.738 & 1.291 & 0.936 & 1.097 & 0.912
    \\
    \hline
    ALFM& 0.891 & 0.735 & 0.870 & 0.758 & 1.280 & 0.976 & 1.065 & 0.870
    \\
    CARP& 0.879 & 0.714 & 0.827 & 0.686 & 1.236 & 0.945 & 1.069 & 0.847
    \\
    RPR& \textbf{0.795} & \textbf{0.652} & \textbf{0.814} & \textbf{0.641} & \textbf{1.141} & \textbf{0.836} & \textbf{0.987} & \textbf{0.784}
    \\
    \hline
    % $\Delta_{DC}$ & 38.2 & 42.5 & 16.5 & 22.3 & 14.2 & 27.5 & 20.9 & 26.2
    % \\
    % $\Delta_{TN}$ & 29.6 & 25.2 & 14.4 & 18.8 & 13.9 & 20.5 & 12.2 & 15.1
    % \\
    % $\Delta_{MP}$ & 13.9 & 24.2 & 7.4 & 13.1 & 11.6 & 10.7 & 10.0 & 14.0
    % \\
    % $\Delta_{AL}$ & 10.8 & 11.3 & 6.4 & 15.4 & 10.9 & 14.3 & 7.3 & 9.9
    % \\
    $\Delta_{CA}$ & 9.6 & 8.7 & 1.6 & 6.6 & 7.7 & 11.5 & 7.7 & 7.4
    \\
    % \hline
    p-value & 9.8E-4 & 5.6E-4 & 1.5E-3 & 4.8E-4 & 1.5E-3 & 1.5E-3 & 9.7E-4 & 4.5E-4
    \\
    \hline
    \hline
    \multirow{2}*{Datasets} & \multicolumn{2}{c|}{Automotive} & \multicolumn{2}{c|}{Toys and Games} & \multicolumn{2}{c|}{Video Games} & \multicolumn{2}{c|}{Yelp}
    \\
    \cline{2-9}
    {} & MSE & MAE & MSE & MAE & MSE & MAE & MSE & MAE
    \\
    \hline
    MF& 2.080 & 1.442 & 1.801 & 1.342 & 1.625 & 1.275 & 1.738 & 1.418
    \\
    FM& 1.599 & 1.265 & 1.197 & 1.094 & 1.656 & 1.287 & 1.726 & 1.414
    \\
    MLP& 1.073 & 1.036 & 1.285 & 1.134 & 1.576 & 1.255 & 1.733 & 1.416
    \\
    NeuMF& 1.023 & 1.011 & 1.210 & 1.101 & 1.568 & 1.252 & 1.682 & 1.397
    \\
    \hline
    D-CON& 1.045 & 0.931 & 1.056 & 0.909 & 1.307 & 1.154 & 1.487 & 1.342
    \\
    T-NET& 0.963 & 0.887 & 1.032 & 0.782 & 1.278 & 1.029 & 1.466 & 1.201
    \\
    MPCN& 0.942 & 0.771 & 1.036 & 0.767 & 1.267 & 1.026 & 1.445 & 1.159
    \\
    \hline
    ALFM& 0.924 & 0.765 & 0.980 & 0.752 & 1.245 & 0.997 & 1.417 & 1.073
    \\
    CARP& 0.913 & 0.712 & 0.943 & 0.735 & 1.238 & 0.978 & 1.406 & 1.079
    \\
    RPR& \textbf{0.896} & \textbf{0.703} & \textbf{0.915} & \textbf{0.727} & \textbf{1.142} & \textbf{0.939} & \textbf{1.303} & \textbf{0.986}
    \\
    \hline
    % $\Delta_{DC}$ & 14.3 & 24.5 & 13.4 & 20.0 & 12.6 & 18.6 & 12.4 & 26.5
    % \\
    % $\Delta_{TN}$ & 7.0 & 20.7 & 11.3 & 7.0 & 10.6 & 8.7 & 11.1 & 17.9
    % \\
    % $\Delta_{MP}$ & 4.9 & 8.8 & 11.7 & 5.2 & 9.9 & 8.5 & 9.8 & 14.9
    % \\
    % $\Delta_{AL}$ & 3.0 & 8.1 & 6.6 & 3.3 & 8.3 & 5.8 & 8.0 & 8.1
    % \\
    $\Delta_{CA}$ & 1.9 & 1.3 & 3.0 & 1.1 & 7.8 & 4.0 & 7.3 & 8.6
    \\
    % \hline
    p-value & 2.6E-3 & 3.5E-3 & 1.1E-3 & 1.4E-3 & 1.6E-3 & 9.5E-4 & 1.0E-3 & 7.8E-4
    \\
    \hline
\end{tabular}
\vspace{-0.5cm}
\end{table*}

%% file: 5_result.tex
\section{Experimental Results}
In order to validate the effectiveness of our proposed method, we conducted extensive quantitative and qualitative experiments to answer the following questions:
\begin{itemize}[leftmargin=*]
\item \textbf{Q1}. Can our proposed method outperform both the state-of-the-art review-involved and traditional recommendation baselines?
% \item \textbf{Q2}. How efficient is RPR as compared to the baseline methods?
\item \textbf{Q2}. How do different components (e.g., the aspect-aware importance weighting component for learning importance offsets) contribute to the overall performance of our proposed model?
\item \textbf{Q3}. How do the key hyper-parameters (e.g., the number of latent factors) affect our model performance?
\item \textbf{Q4}. Can our model provide explicit interpretations? %\item \textbf{Q5}.
%\item \textbf{Q6}. Can we explicitly verify the mapping relationship between the user-preferred and user-rejected aspects?
\end{itemize}

\subsection{Q1: Performance Comparison}
The results of our method and other baselines over the experimented datasets are presented in Table \ref{tab:q1}. The key observations can be summarized as follows:
\begin{itemize}[leftmargin=*]
\item The recommendation methods from the first category obtain the worst performance on all datasets. And the deep learning-based methods (i.e., NeuMF and MLP)  can achieve superior performance compared to the factorization-based ones (i.e., FM and MF), since they can model more complex interactions than the factorization ones.

\item The review-involved methods consistently surpass the interaction-based ones, demonstrating that reviews contain valuable side information for accurate recommendation. Moreover, amongst the review-involved baselines, it is obvious that MPCN largely outperforms both DeepCoNN (D-CON) and TransNet (T-NET), which mainly benefits from the fact that MPCN adopts the pointer-based scheme to filter important reviews for recommendation. Nevertheless, TransNet outperforms DeepCoNN, since it takes the review of target user-item pair as the approximation object in the training process.
%, distinguishing the reviews to a certain extent.

\item The aspect-aware methods surpass the plain review-involved ones in most cases. This indicates that capturing aspect-aware information from reviews is effective. Among these baselines, the recently proposed CARP model yields a better result than ALFM. It is because that CARP introduces the capsule network, which is proposed to model the complex relations among features, while ALFM employs a general aspect-aware topic model.

\item Finally, we can observe that RPR substantially outperforms all the baselines on the eight datasets. When comparing with the review-involved baselines regarding the MSE metric, its relative improvement is satisfying with gains up to $38.2\%$ (DeepCoNN), $29.6\%$ (TansNet), and $13.9\%$ (MPCN), respectively. Moreover, it is obvious that RPR consistently exceeds the aspect-aware baselines by a considerable margin. Jointly analyzing Fig.~\ref{fig:imbalance} and Table~\ref{tab:q1}, we observed that the improvements over the best baseline are more significant in datasets \textit{Digital Music}, \textit{Tools Improvement}, \textit{Video Games}, and \textit{Yelp}, where the negative reviews are more sufficient.  This observation indicates that discriminately treating reviews with different polarities would promote the recommendation performance.
% Moreover, we also discussed the efficiency of RPR and other methods in the supplementary material.
\end{itemize}

\subsection{Q2: Ablation Study}
We conducted detailed ablation studies to validate how each component contributes to the overall performance of our model.
In particular, we compared our proposed model with the following variants:
\begin{itemize}[leftmargin=*]
\item \textbf{Base:} The base refers to our complete model with the optimal setting.
\item \textbf{Coarse-grained Model:}
Instead of extracting the word-wise aspect importance distribution in Eqn.($\ref{eq:cnn}$), this variant introduces the max-pooling layer to the convolutional layer in the base.
% and directly learns the importance of user-preferred and user-rejected aspects for the user from the positive and the negative reviews, respectively.
\item \textbf{W/o Review Polarity:}
We removed the components of distinguishing the positive and negative reviews of the user, and followed the previous review-involved methods to collect all the reviews as the user document.
% Similarly, the ability of distinguishing the user-preferred and user-rejected aspects of items is removed.
% The same extractor and score prediction component are employed for the final rating prediction.
\item \textbf{Uniform Aspect Importance:}
We replaced $\bm{\rho}^{p+}_u$ and $\bm{\rho}^{r+}_u$ in Eqn.($\ref{eq:objective}$) with the uniform importance distributions, i.e., all user-preferred and user-rejected aspects are assumed to be equally important.
\item \textbf{W/o Importance Offset:}
We removed the aspect-aware importance weighting component from the complete model. It is worth noting that the imbalance problem is not specifically tackled in this variant.
\end{itemize}

\begin{table}[t]\centering
\caption{Performance comparison of the model variants on the \emph{Automotive} and \emph{Video Games} datasets.}
\label{tab:ablation}
\vspace{-0.3cm}
\begin{tabular}{ccc}
\hline
Setup & Automotive & Video Games
\\
\hline
Base & 0.896 & 1.142
\\
Coarse-grained Model & 0.952 & 1.319
\\
W/o Review Polarity & 0.949 & 1.325
\\
Uniform Aspect Importance & 0.964 & 1.330
\\
W/o Importance Offset & 0.947 & 1.384
\\
\hline
\end{tabular}
\vspace{-0.4cm}
\end{table}

Table \ref{tab:ablation} shows the MSE results of the above variants on the \emph{Automotive} and \emph{Video Games} datasets.
First, the word-wise extraction of aspect importance outperforms the document-wise one,
which verifies that the fine-grained semantic extraction boosts the performance of review-involved recommendation.
Second, we can observe that ignoring the polarities of reviews will deteriorate the performance, it is thus necessary to distinguish the positive and negative reviews for review-involved recommendation.
Additionally, it is reasonable for a user to attach different importance to different aspects of an item. Thus, uniforming the aspect importance would limit the modeling capacity of user preferences.
Finally, the result of the last variant reflects that the aspect-aware importance weighting module in our model can effectively reduce the impact
of the review imbalance.

\subsection{Q3: Effectiveness of Key Hyper-parameters}
\begin{figure}[tbp]
\centering
\includegraphics[width=0.75\linewidth]{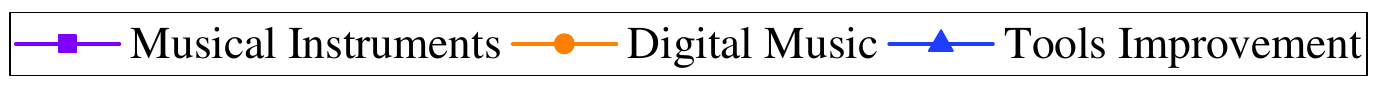}
\\
%\subfigure{
\begin{minipage}[t]{0.45\linewidth}
\centering
\includegraphics[width=\linewidth]{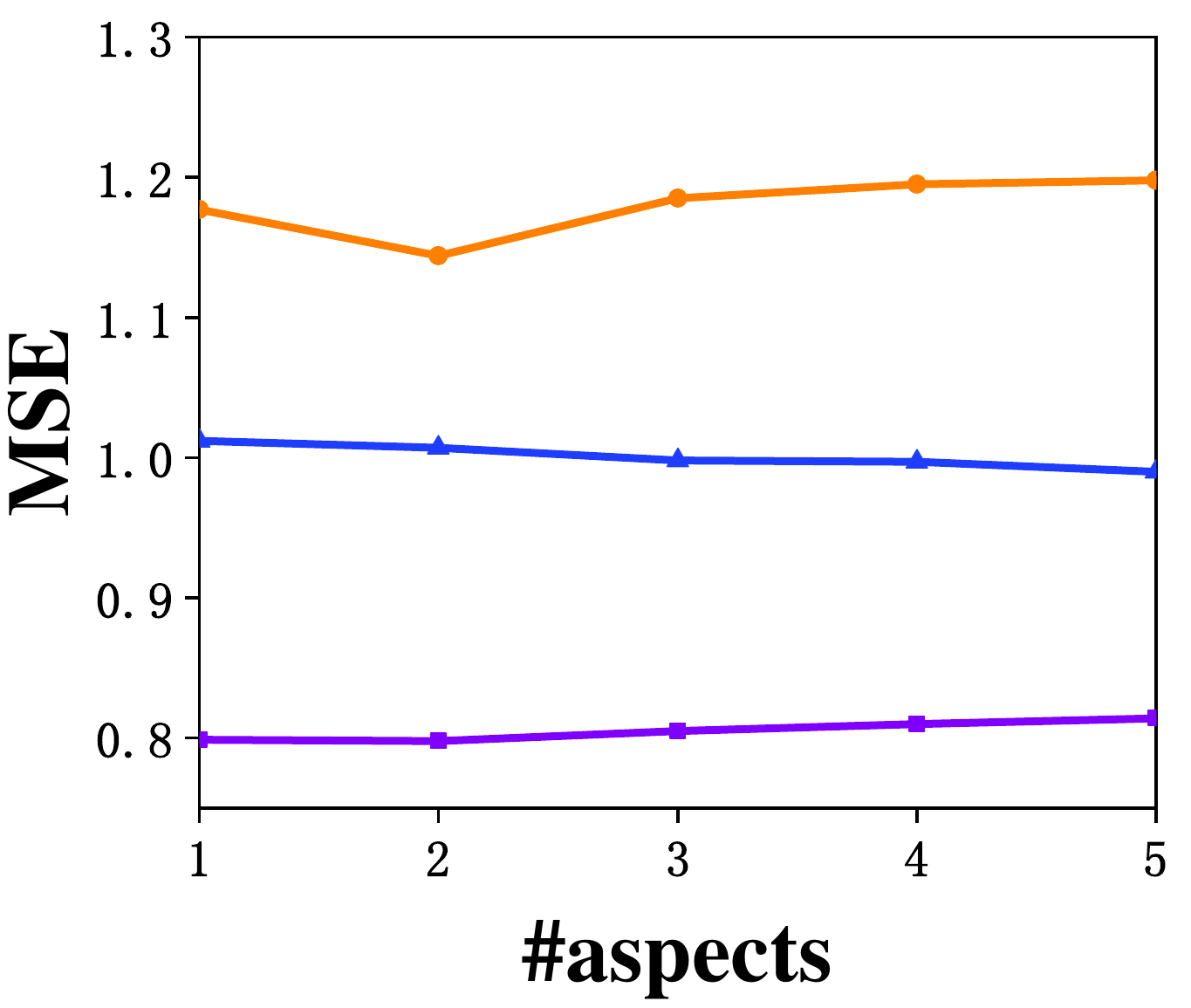}
%\caption{fig1}
\end{minipage}%
%}%
%\subfigure{
\begin{minipage}[t]{0.45\linewidth}
\centering
\includegraphics[width=\linewidth]{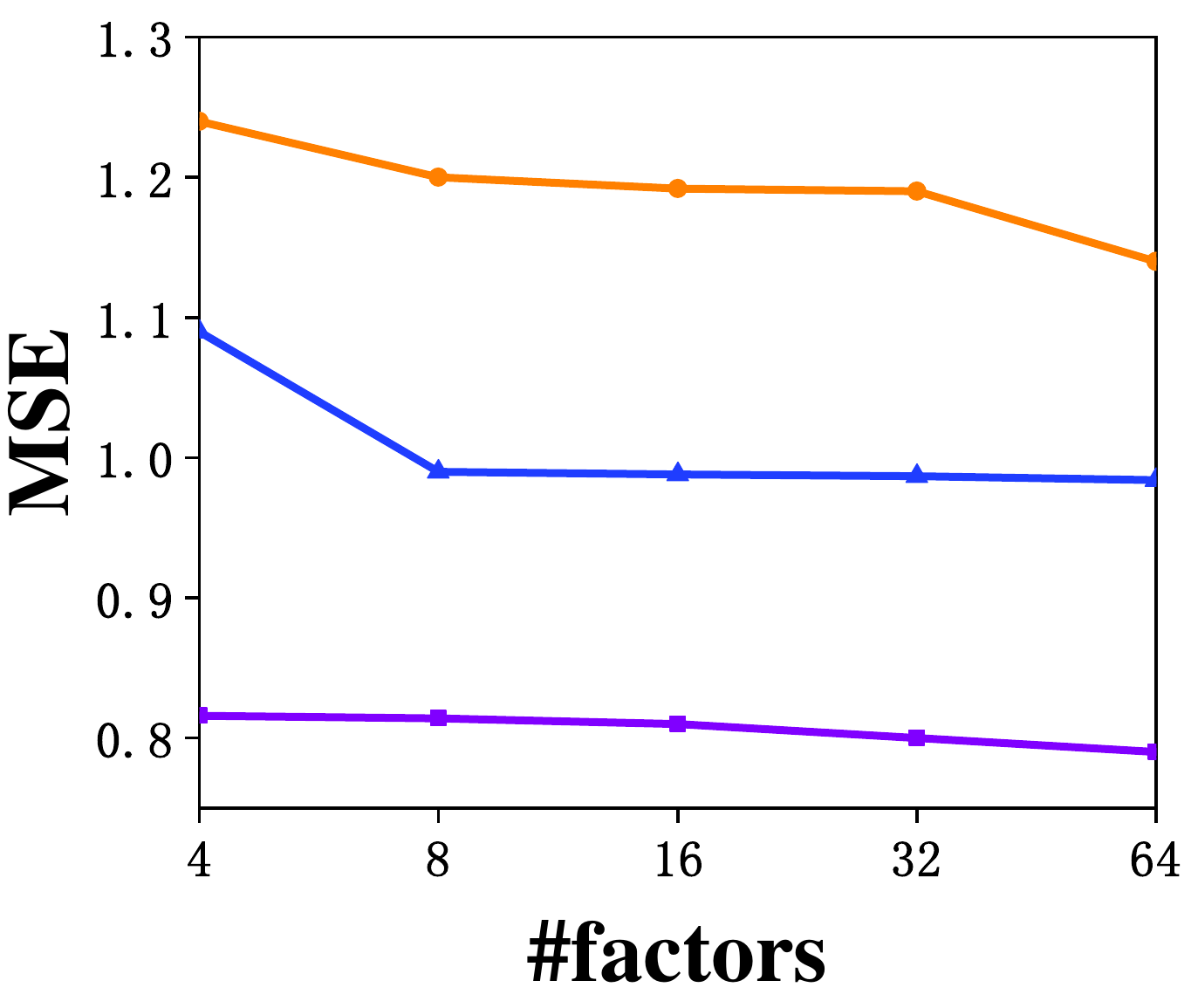}
%\caption{fig2}
\end{minipage}%
\\
\begin{minipage}[t]{0.45\linewidth}
\centering
\includegraphics[width=\linewidth]{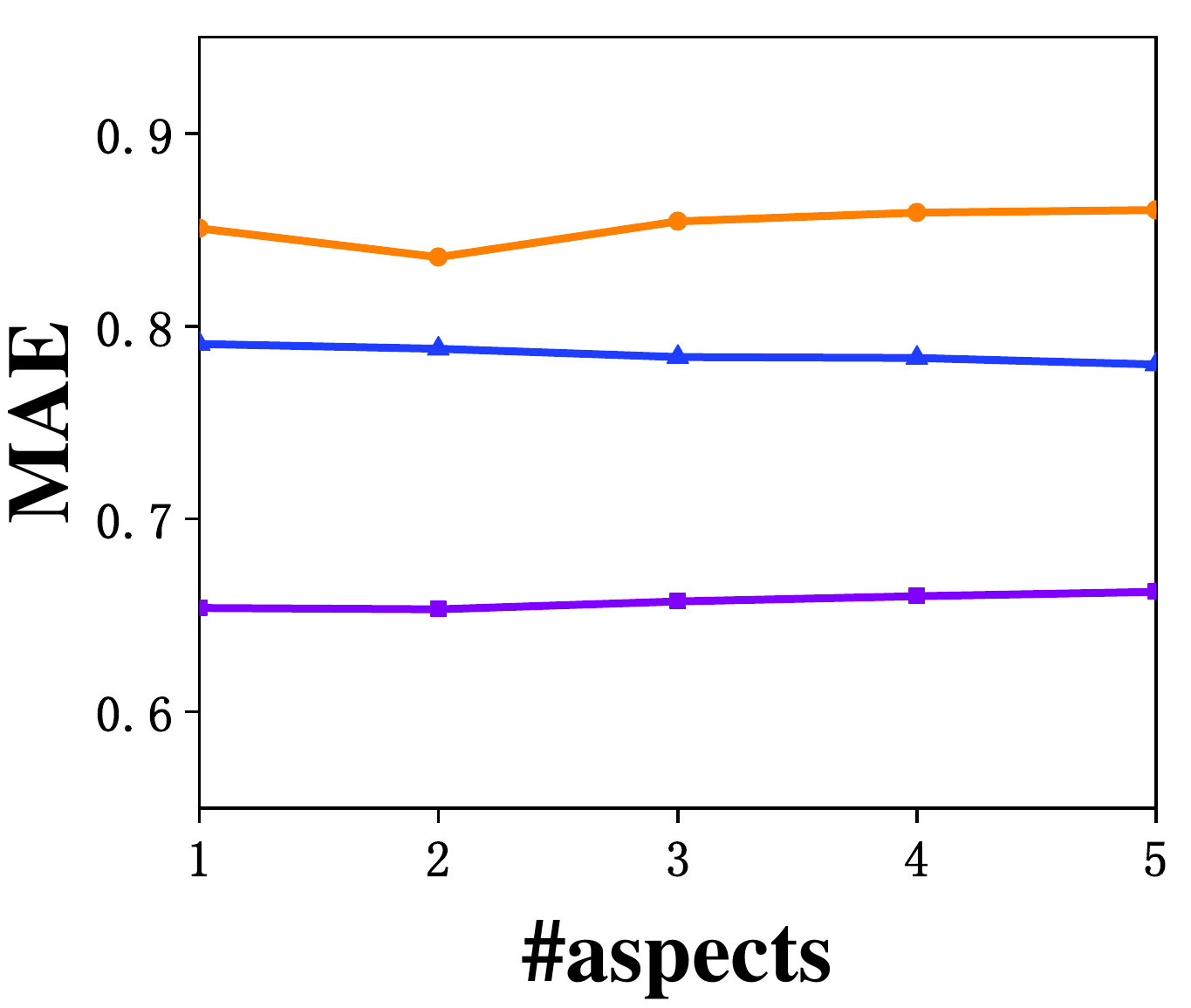}
%\caption{fig1}
\end{minipage}%
%}%
%\subfigure{
\begin{minipage}[t]{0.45\linewidth}
\centering
\includegraphics[width=\linewidth]{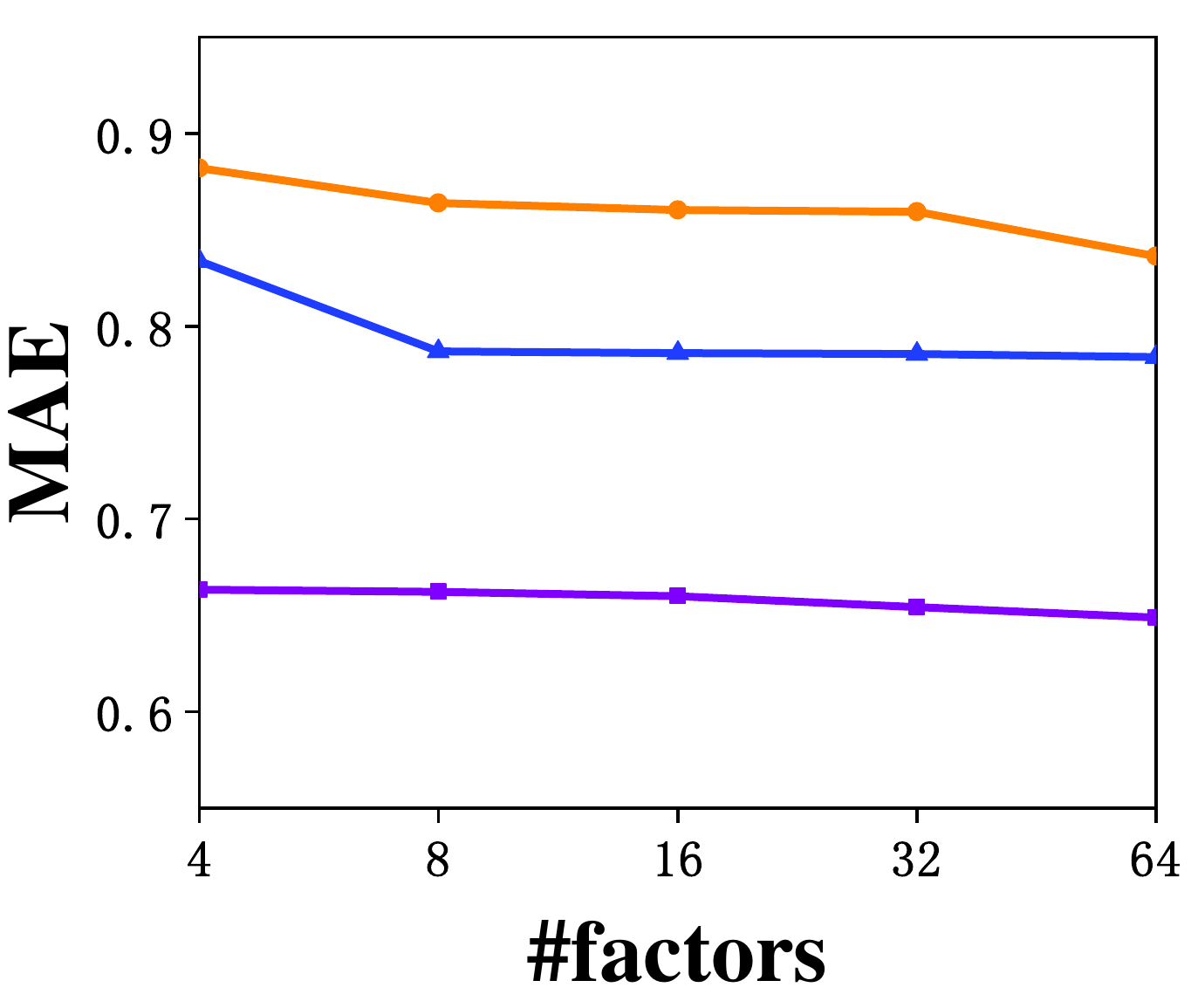}
%\caption{fig2}
\end{minipage}%
%}%
\centering
%\vspace{-0.5cm}
\caption{Effect of the number of aspects and factors in our model.}
\label{fig:effect}
\vspace{-0.3cm}
\end{figure}

In this subsection, we analyzed the effectiveness of the key hyper-parameters in our method for the overall performance.
We primarily focused on the number of aspects and latent factors
(i.e., the dimension of embeddings $\mathbf{p}_u$ and $\mathbf{q}_i$). Fig.~\ref{fig:effect} shows the performance variations
with changing the number of aspects and latent factors on three datasets.

\textbf{Number of Aspects.}
We set the number of latent factors to $32$ for better studying the effect of aspects. With the number of aspects changing from 1 to 5 as illustrated in Fig.~\ref{fig:effect}, we observed that the optimal number of aspects varies with different datasets,
which is probably because that users comment different aspects in reviews for different categories of items.
%Besides, the optimal aspect numbers increase with the scale expansion of datasets, and this demonstrates that plenty of
%reviews help the model to characterize more aspects.
Additionally, promising performance can be obtained when the number of user-preferred/user-rejected aspects is from 2 to 4.
%In general, we can conclude varying the number of user-preferred/user-rejected aspects has little impact on the overall model performance.

\textbf{Number of Latent Factors.}
%In this part of experiments, we fix the number of items' good and bad aspects to $6$, in order to investigate the influence of the number of the latent factors.
To study the effect of latent factors, we fixed the number of user-preferred/user-rejected aspects to $3$.
From the Fig.~\ref{fig:effect}, it can be seen that the MSE decreases with increasing the latent factors, since the rating prediction is still based on matrix factorization in our model. Therefore, increasing latent factors can better represent the user/item, contributing to a more accurate rating prediction.
% It is worth noting that we also discussed the joint effectiveness of aspects and factors for a better understanding in the supplementary material.

\begin{figure}[tbp]
\centering
%\includegraphics[width=0.8\linewidth]{list.pdf}
%\\
%\subfigure[\textbf{Office Product}]{
\begin{minipage}[t]{0.495\linewidth}
\centering
\includegraphics[width=\linewidth]{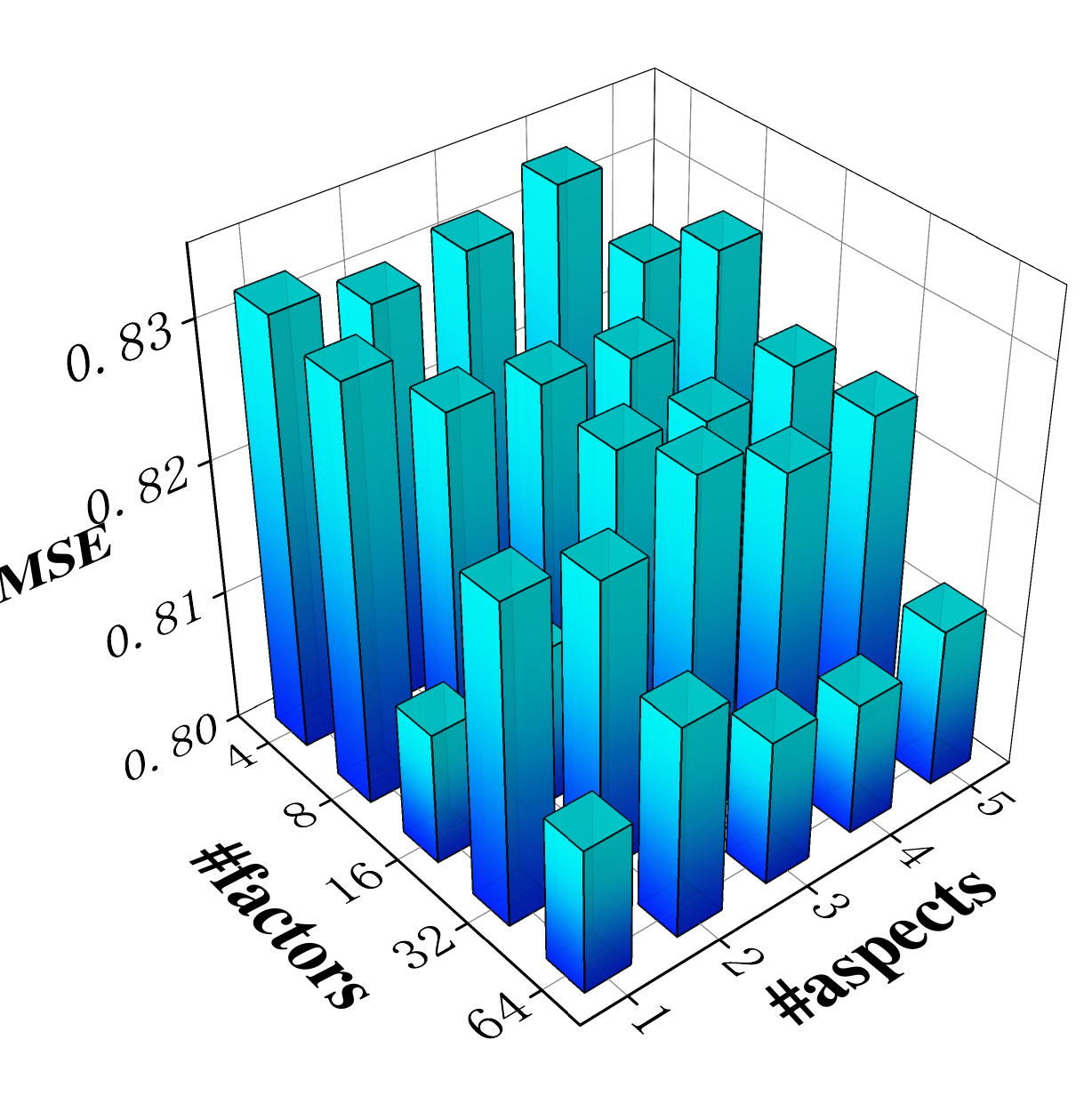}
%\caption{fig1}
\centerline{\footnotesize Office Product}
\end{minipage}%
%}%
%\subfigure[\texbf{Yelp}]{
\begin{minipage}[t]{0.495\linewidth}
\centering
\includegraphics[width=\linewidth]{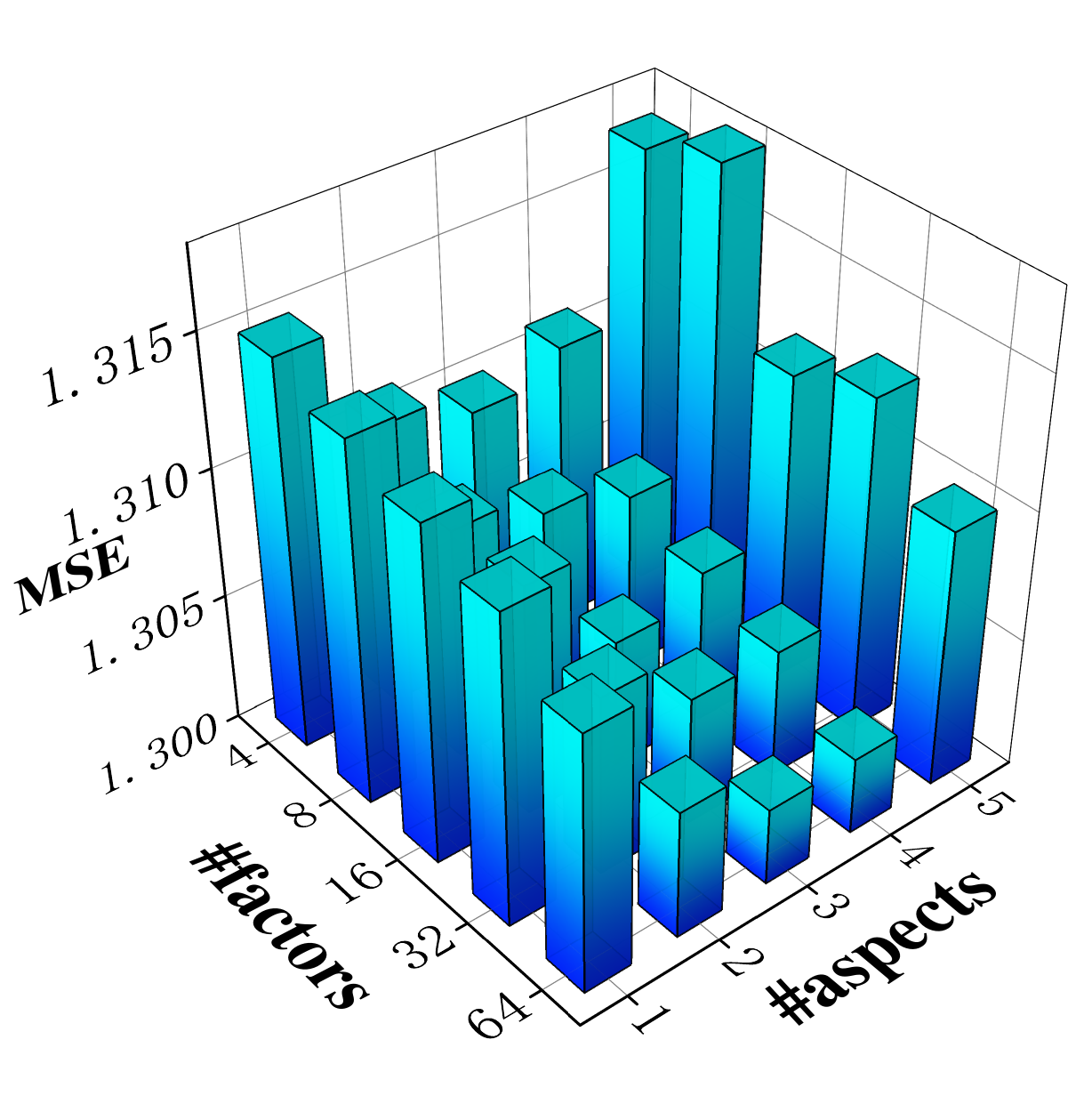}
%\caption{fig2}
\centerline{\footnotesize Yelp}
\end{minipage}%
%}
\\
\begin{minipage}[t]{0.495\linewidth}
\centering
\includegraphics[width=\linewidth]{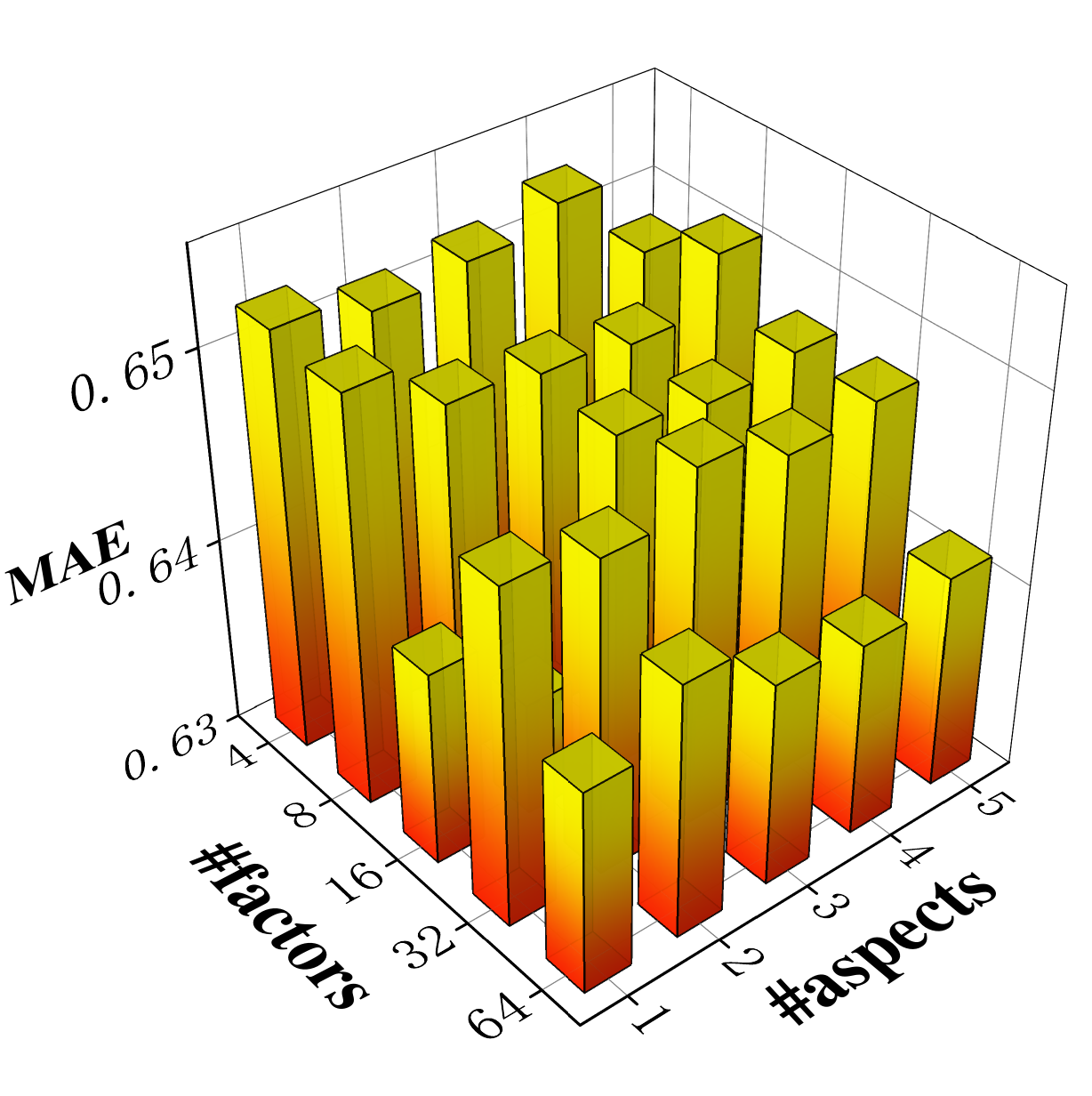}
%\caption{fig1}
\centerline{\footnotesize Office Product}
\end{minipage}%
%}%
%\subfigure[\texbf{Yelp}]{
\begin{minipage}[t]{0.495\linewidth}
\centering
\includegraphics[width=\linewidth]{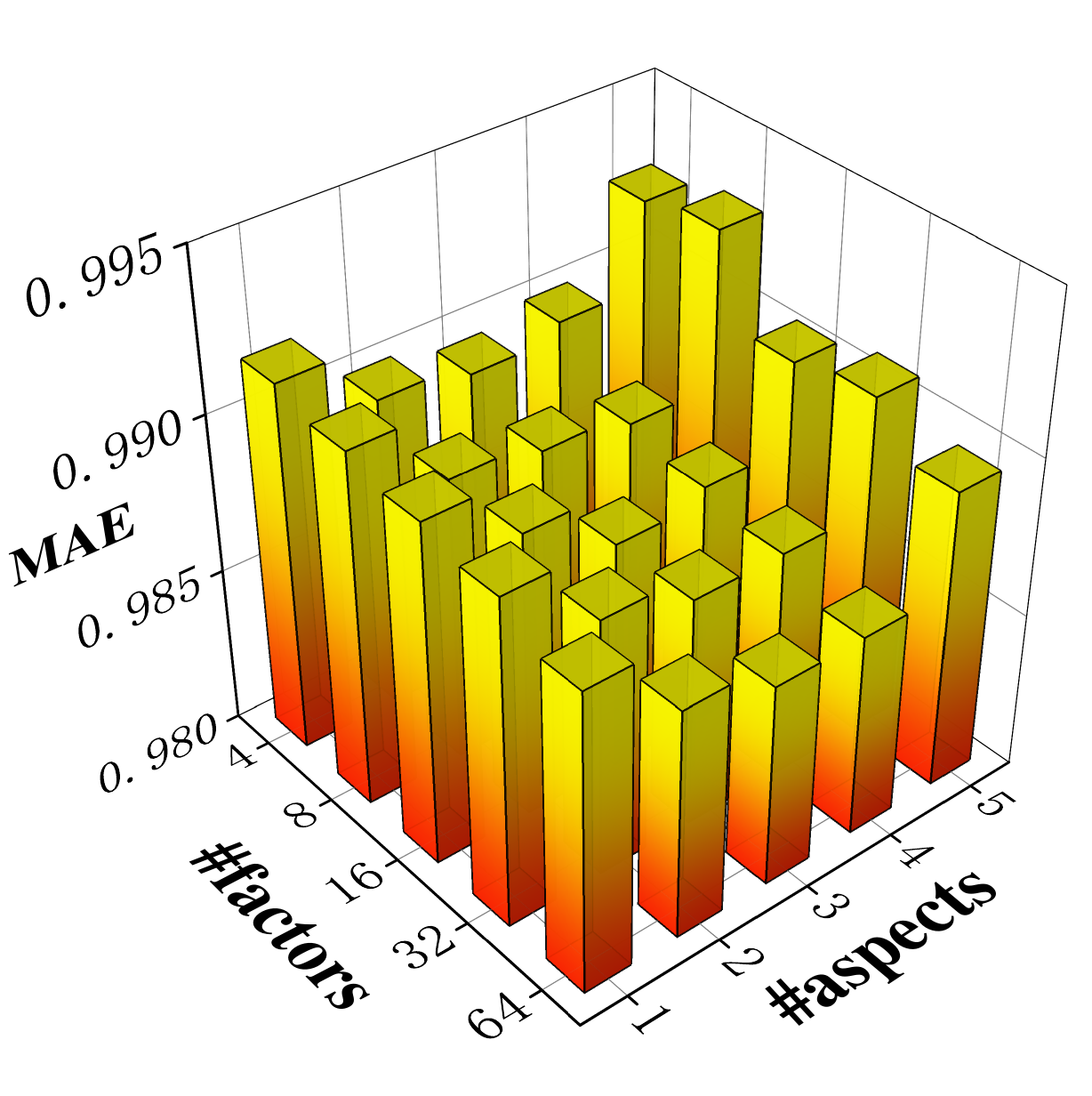}
%\caption{fig2}
\centerline{\footnotesize Yelp}
\end{minipage}%
\centering
%\vspace{-0.4cm}
\caption{Confounding effect of the number of aspects and factors.}
\label{fig:effects}
\vspace{-0.6cm}
\end{figure}

To visualize the joint effects of aspects and factors, we presented 3D figures by varying the number of user-preferred/user-rejected aspects in $\{1, 2, 3, 4, 5\}$, and the number of factors in $\{4, 8, 16, 32, 64\}$ and illustrated the results of \emph{Office Products} and \emph{Yelp} datasets. From Fig.~\ref{fig:effects}, it can be  recognized that the optimal numbers of aspects and factors are different across datasets. In general, more latent factors usually lead to better performance, while the optimal number of user-preferred/user-rejected aspects might depend on the review details of different datasets.

\begin{table*}\centering
\caption{Positive and negative reviews of a randomly selected user from \emph{\textbf{Musical Instruments}}.}
\label{tab:reviews}
% \vspace{-0.3cm}
\setlength{\tabcolsep}{1.0mm}{
\begin{tabular}{|p{0.45\linewidth}<{\leftline}|p{0.45\linewidth}<{\leftline}|}
    \hline
    Positive Reviews & Negative Reviews
    \\
    \hline
    ...My guitar player had a different lock system, and his \$1000 Les Paul fell to the stage, completely knocking it out of tune.
    Mine stayed locked perfectly...
    &
    It's a decent unit, but stopped working completely after about 6 months. Tried every thing I could, but it's gone.
    I would not recommend this pedal.
    \\
    \hline
    These strings are nice and easy to fly over. No buildup, or residue on your fingers either. I really like this stuff.
    &
    The plastic piece that screws the wire into the end was made of very thin plastic, and cracked in two within a week
    of light use.
    \\
    \hline
    Nothing sounds as bright as these strings. I've tried many, and these are the best by far. I know, cause I play guitar
    really god-like.
    &
    It stopped working after 2 gigs. I'm not sure why, but it is very frustrating. I guess you get what you pay for here.
    \\
    \hline
    I often use this  to record my band's gigs. Now if they would make one that will hold a PAR 38 so I can turn
    my extra microphone stands into single can lighting racks!
    &
    I expected more from this manufacturer, but I guess the quality is not the same as it used to be.
    \\
    \hline
\end{tabular}}
\vspace{-0.5cm}
\end{table*}

\subsection{Q4: Interpretability}
In our RPR model, a user's preference on an item depends on the scores and the importance of both user-preferred and user-rejected aspects to the user.
The importance on user-preferred/user-rejected aspects is computed by the aspect-specific semantic embeddings in the user's positive/negative reviews.
Since we assume one element corresponds to one aspect, accordingly, review words which hold the largest corresponding elements in their semantic embeddings are adopted as the semantic explanation for this aspect.
Table~\ref{tab:reviews} records the positive and negative reviews of a randomly selected user from the \emph{Musical Instruments} dataset.
In our experiments, we set the number of user-preferred and user-rejected aspects to $2$, and filtered their corresponding review words.
Table~\ref{tab:word} shows the top 10 aspect words (we removed stop words for better illustration).
Based on this experiment, the two user-preferred aspects can be semantically interpreted as ``Performance'' and ``Fine Strings'',
and the two user-rejected aspects can be interpreted as ``Poor Quality'' and ``Crack Sensitive''.

Next, we aimed to study the interpretability of our proposed model on high and low ratings~\cite{wang2018explainable, wang2018tem}.
From the same dataset, we chose ``item 1'' and ``item 2'', which are rated with 5 and 1 by the selected user, respectively.
We firstly obtained the selected user's aspect importance on the user-preferred and user-rejected aspects (i.e., $\bm{\rho}^{p+}_u$ and $\bm{\rho}^{r+}_u$), and then computed the aspect scores on the user-preferred and user-rejected aspects (i.e., $\mathbf{s}_{u,i}^{p}$ and $\mathbf{s}_{u,i}^{r}$) of ``item 1'' and ``item 2'', respectively.
As shown in Table \ref{tab:imp}, we could observe that the selected user pays more attention on the user-preferred aspect ``Fine Strings''
and the user-rejected aspect ``Poor Quality''.
On the user-preferred aspects, we could see that ``item 1'' is a better match to the user's preference compared with ``item 2'', because the user-preferred aspect scores
of ``item 1'' are higher.
On the user-rejected aspects, we could observe that the scores of ``item 1'' are both negative values.
This probably means that ``item 1'' has none of these user-rejected aspects.
% , even opposite (e.g., be superior in quality).
The scores of ``item 2'' on user-rejected aspects are positive values, indicating that ``item 2'' possesses these user-rejected aspects
in the user's opinion.
This example illustrates that our proposed model is capable of providing the interpretability for recommendation
from the view of the user-preferred and user-rejected aspects.

%% file: 6_conclusion.tex
\section{Conclusion and Future Work}
In this paper, we present a review polarity-wise recommender, dubbed as RPR, which treats reviews with different polarities discriminately. Specifically, RPR simultaneously learns the scores and importance of the user-preferred and user-rejected aspects to a user. The final rating is then estimated via the mathematical difference of the positive and negative scores, which are the weighted sum of the relevance scores with the corresponding importance.
%RPR is able to extract fine-grained semantical information about the user's aspect importance covered in each review word. Besides, RPR can capture a user's positive and negative attitudes towards an item for a more convincing and personalized rating prediction.
To overcome the problem of imbalanced review polarity, RPR implements an aspect-aware importance weighting module to effectively learn the mapping relations for one aspect importance based on the other.
In addition to its remarkable performance over eight datasets, RPR is also capable of providing explicit explanation for the recommendation results.
% In the experiments on eight real-world datasets, RPR outperforms the state-of-the-art methods on the rating prediction task.
% To demonstrate the effectiveness of each component in RPR, corresponding experiments and studies have been conducted.

In future, we will apply pairwise learners to strengthen RPR and validate its effectiveness via extensive experiments. Moreover, we are particularly interested in discriminating the user-preferred and user-rejected aspects of multi-media items~\cite{mmgcn}, which contain abundant aspect information to reflect users' preferences.
% To build a multi-media recommender system, we plan to develop an effective method to extract user-preferred and user-rejected aspects from multi-view and multi-modal data~\cite{he2014comment, wang2012multimodal}.
Another interesting direction is to extend RPR to solve the long-tail recommendation problem by extracting semantic information from the attribute descriptions of less frequent items, which requires urgent solution for e-commerce platforms.
% While our current model primarily focuses on the personalized recommendation for individuals, it is interesting to develop PRP for groups of users, which can help the decision-making for social groups.
\begin{table}[tbp]\centering
\caption{Top 10 words for each user-preferred and user-rejected aspect of the selected user from \emph{\textbf{Musical Instruments}}.
The ``aspect labels'' are attached based on the interpretation of the aspect.}
\label{tab:word}
\vspace{-0.3cm}
\begin{tabular}{cc|cc}
    \hline
    \multicolumn{2}{c|}{User-preferred Aspects} & \multicolumn{2}{c}{User-rejected Aspects}
    \\
    \hline
    Performance & Fine Strings & Poor Quality & Crack Sensitive
    \\
    \hline
    record & tried & but & cracked
    \\
    system & buildup & working & lasted
    \\
    gigs & fingers & stopped & unit
    \\
    stayed & extra & use & months
    \\
    tune & perfectly & quality & tape
    \\
    band & sounds & frustrating & piece
    \\
    guitar & strings & wire & gigs
    \\
    stuff & easy & end & pedal
    \\
    stage & bright & manufacturer & work
    \\
    knocking & guitar & pay & guess
    \\
    microphone & lock & decent & screw
    \\
    \hline
\end{tabular}
\vspace{-0.4cm}
\end{table}
\begin{table}[tp]\centering
\caption{Interpretation on why the user rated ``item 1'' and ``item 2'' with 5 and 1, respectively. The examples are obtained based on \emph{\textbf{Musical Instruments}}.}
\label{tab:imp}
\vspace{-0.3cm}
\begin{tabular}{c|ccc}
\hline
Aspects & Importance & Score (1) & Score (2)
\\
\hline
Performance & 0.992 & 0.71 & 0.681
\\
Fine Strings & 1.873 & 0.83 & 0.752
\\
\hline
Poor Quality & 1.996 & -0.67 & 0.185
\\
Crack Sensitive & 0.967 & -0.898 & 0.723
\\
\hline
\end{tabular}
\vspace{-0.3cm}
\end{table}